\documentclass[twocolumn,pra,amsmath,amssymb,superscriptaddress,notitlepage,footinbib]{revtex4-1}

\usepackage{braket}
\usepackage{amsmath}
\usepackage{amsfonts}
\usepackage{xcolor}
\usepackage{amssymb}
\usepackage{graphicx}
\usepackage[caption=false]{subfig}
\usepackage{natbib}

\begin{document}
\title{Electron Counting Statistics for Non-Additive Environments}
\author{Conor McConnell}
\email{conor.mcconnell@postgrad.manchester.ac.uk}
\affiliation{School of Physics and Astronomy, University of Manchester, Oxford Road, Manchester, M13 9PL}
\author{Ahsan Nazir}\email{ahsan.nazir@manchester.ac.uk}
\affiliation{School of Physics and Astronomy, University of Manchester, Oxford Road, Manchester, M13 9PL}
\date{\today}

\begin{abstract}
Molecular electronics is a rapidly developing field focused on 
using
molecules 
as the structural basis for electronic components. 
It is common in such devices 
for the system of interest to couple simultaneously to multiple environments. Here we consider a model comprised of a double quantum dot (or molecule) coupled strongly to vibrations and weakly to two electronic leads held at arbitrary bias voltage. The strong vibrational coupling invalidates treating 
the bosonic and electronic environments simply as acting additively, as would be the case 
in the weak coupling regime or for flat leads at infinite bias. Instead, making use of the reaction coordinate framework we incorporate the dominant vibrational coupling effects within an enlarged system Hamiltonian. This allows us to derive a non-additive form for the lead couplings that accounts properly for the influence of strong and non-Markovian coupling between the double dot system and the vibrations. Applying counting statistics techniques we track electron flow between the double dot and the electronic leads, revealing both strong-coupling and non-additive effects in the electron current, noise and Fano factor. 
\end{abstract}

\maketitle
\section{Introduction}
The original idea for a molecular nanojunction was put forward by Aviram and Ratner in 1974 \cite{Aviram74}, in which they suggested using a single organic molecule to develop a simple electronic rectifier. Since this initial proposal 
the field of molecular electronics has grown quickly, 
both in 
the  
construction of electronic components from molecular building blocks, and in the modelling and measurement of devices to tailor and probe their useful characteristics \cite{Tao06, Sun14, Aradhya13, Chen07, Nichols15}. 
The past few years have seen numerous breakthroughs in the production of molecular nanojunctions~[\onlinecite{Reed252,Dekker97,Park99,Kubatkin03,Cui01,Nicewarner01,Nichols15,Sun14}], 
while theoretical advances have also been made 
with recent works examining a wide variety of nanojunction attributes; examples include current-voltage \cite{Tao06}, conductance \cite{Nitzan03,Chen07} and noise (Fano factor) \cite{Koch05} characteristics, along with thermoelectric, optical and electromechanical processes \cite{Aradhya13,Miao17, Segal16, Segal19}. One aspect of key importance in terms of further experimental progress and the viability of applications is understanding the role of electron-vibrational coupling~\cite{Kushmerick04, Koch05, Koch205, Galperin06, Brandes07, Brandes09, Thoss09, Thoss11, Emary12, Santamore13, Pigeon17, Stones17, Sowa17, Sowa217}. 
{
This has been explored, for example, through nanojunction models with coupling 
to 
individual vibrational modes~\cite{Koch05,Mukamel06,Santamore13,Schaller14,Jiang17}}.  
{Our focus here 
is on 
a continuum of modes, where 
we wish to avoid 
either a weak electron-vibrational coupling approximation 
or the infinite lead bias regime, 
both of which can be used to simplify the resulting formalism. 
We thus aim to present 
an analysis 
of nanojunction models, and specifically electron counting statistics techniques, 
within arbitrary bias regimes for strong and non-Markovian vibrational coupling 
\cite{Strasberg16,Sowa217,Sowa17,Krause15}.} 

We shall consider a molecular double quantum dot (DQD) system comprised of two electronic levels, which are coupled to a pair of fermionic leads and 
a bosonic vibrational (phonon) environment. This set up is a working model for a molecular nanojunction in which vibrational influences are significant. 
We address the challenge of strong electron-phonon coupling using the reaction coordinate (RC) mapping \cite{IlesSmith14,IlesSmith16,Strasberg16,Maguire18}. 
{This is a (unitary) Hamiltonian level mapping that transforms the coupling of the DQD electronic system to the original continuum phonon environment to coupling only to a single collective mode (the RC), which in turn interacts weakly with a residual bosonic environment. It thus allows for the inclusion of the RC 
within an enlarged (augmented) system along with
the DQD to capture the dominant strong-coupling and non-Markovian effects.} 
We then treat the remaining (residual) bosonic environment and the fermionic leads as being weakly coupled to the augmented system 
using standard Born-Markov (BM) master equation methods. 

As a means to compare with common approximations, we shall analyse 
our 
model within three 
approaches. We begin with a standard weak coupling theory, in which we treat the interactions between our system and each of the environments using the BM approximations without any RC mapping. We shall compare 
this weak coupling approach with 
two strong-coupling methods {applied to the same underlying Hamiltonian. In the first of these we consider the phonons within the RC mapped formalism to incorporate strong electron-phonon interactions.} However, the leads will be treated through standard dissipators having no indication of this strong coupling, i.e.~they will be incorporated additively. In the second strong coupling approach we again consider the phonon bath within the RC formalism, however, unlike before, we shall then consistently derive the lead dissipators as being weakly coupled to our enlarged augmented system {that includes both the DQD and the RC}. As such the leads become ``aware" of the phonon coupling within the augmented system, resulting in a non-additive treatment. 
In this way we can build up our understanding of the system from the weak coupling theory, to a strong coupling additive theory, and then finally arrive at a strong coupling theory containing non-additive contributions. This allows us to highlight and isolate both strong-coupling and non-additive vibrational contributions to the system transport properties. 

We shall employ full counting statistics techniques to probe electron transport, which enables 
calculation of the zero frequency current, noise and Fano factor of our DQD system \cite{Flindt05,Flindt10}. From these 
quantities we can ascertain information regarding the electron correlations present within the probability distribution of our system. Originally, these techniques were 
used alongside perturbative treatments for the phonon environment \cite{Brandes07,Schaller14}. However 
recent works 
have successfully applied 
electron counting statistics to systems where the electron-phonon interaction is non-perturbative, for example  
through the hierarchical equations of motion approach \cite{Stones17, Cerrillo16} or a polaron transformation \cite{Sowa17}.  
One feature of the RC treatment worth noting is that although our electronic system dynamics is non-Markovian, we work with a Markovian master equation in the augmented (RC mapped) representation, and 
as such the electron counting statistics formalism is easily applied. 

Molecular nanojunction models are also commonly probed within the infinite bias regime, 
whereby the leads are held at chemical potentials that ensure a unidirectional flow of electrons. 
For flat lead spectral densities the infinite bias case is necessarily additive (see also below), however the RC formalism allows us to 
relax this boundary, thus providing insight into finite bias regimes and the impact of non-additive lead-phonon contributions on the transport statistics. 

The paper is organised as follows. In Section \ref{Model} we shall introduce 
the theoretical model, 
briefly commenting on each component of the Hamiltonian, before outlining the derivation of the weak-coupling treatment in Section~\ref{wcme}. In Section \ref{RCMap} we shall discuss the implementation of the RC formalism for strong phonon coupling and derive the lead components for the 
non-additive master equation. Section \ref{FCS} describes the procedure for implementing full counting statistics within our system. In Section \ref{Results} we shall present our results. We demonstrate that non-additive contributions have an impact at finite bias when considering the zero frequency current, noise, and Fano factor for flat leads, and even also at infinite bias if the leads are considered to have Lorentzian profiles. Finally, in Section \ref{Conc} we shall summarise our findings and discuss potential further work.

\section{Theoretical Model}\label{Model}

\begin{figure}[t]
\center
 \includegraphics[width=0.8\linewidth]{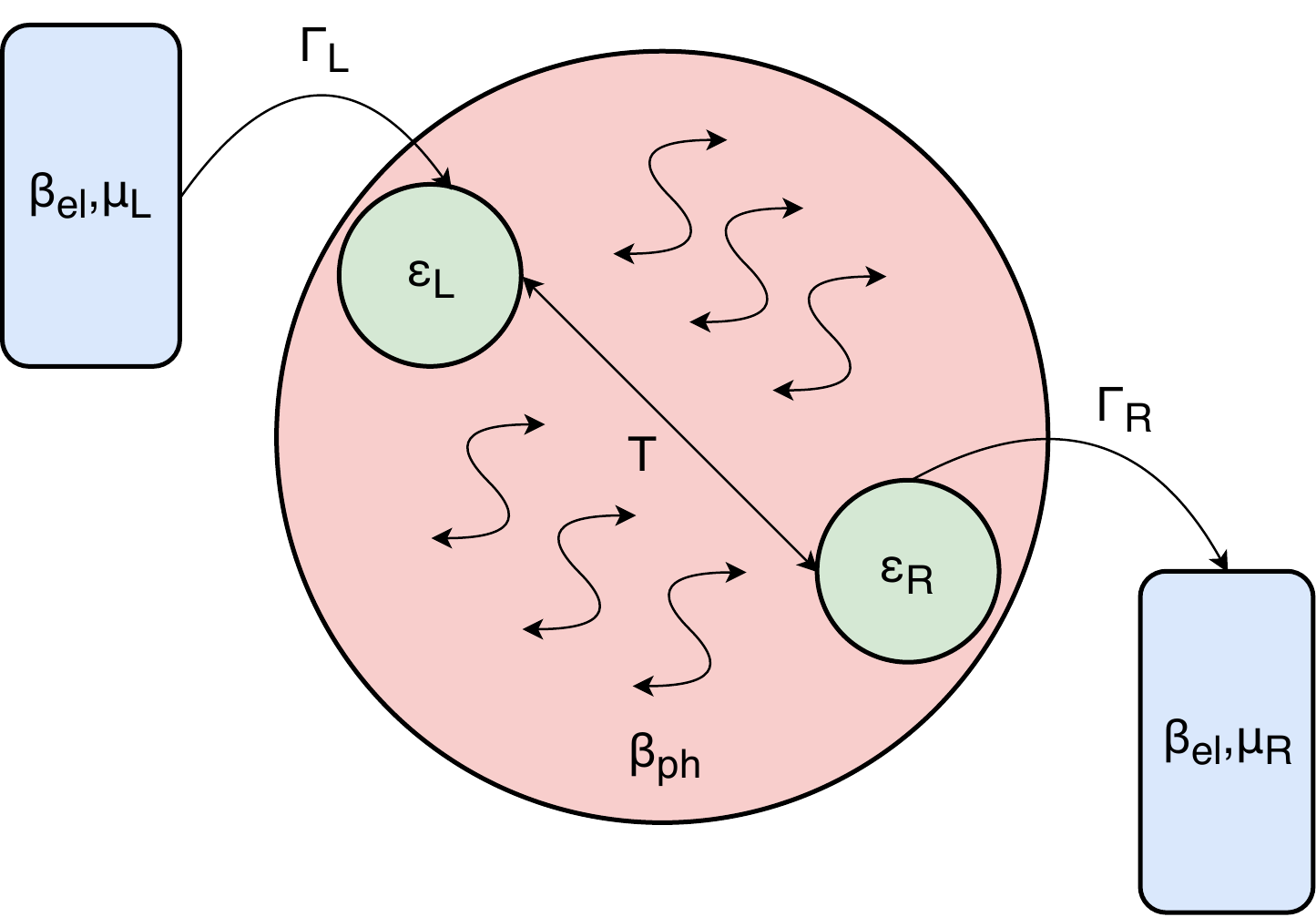}
\caption{Schematic diagram of a double quantum dot system (energies $\epsilon_{\rm L/R}$ and tunnelling element $T$) coupled collectively to a phonon bath at inverse temperature $\beta_{\rm ph}$, with each individual dot coupled through $\Gamma_{\rm L/R}$ to a respective fermionic lead at inverse temperature $\beta_{\rm el}$ and chemical potential $\mu_{\rm L/R}$.}
\label{diagram1}
\end{figure}

A schematic representation of our model system is given in Fig.~\ref{diagram1}. We consider a DQD that is coupled to a 
phonon environment and two fermionic leads. Each dot is represented by an electronic level 
with energies $\epsilon_{\rm L}$ and $\epsilon_{\rm R}$ for the left and right dot, respectively. 
Throughout this work we impose the Coulomb blockade limit, ensuring that only one electron is present on either dot at any given time. 
We thus work in a reduced system space spanned by $\{\ket{0},\ket{L},\ket{R}\}$, which we term the site basis, where $\ket{0}$ represents the case in which neither dot is occupied and $\ket{L}$ ($\ket{R}$) 
denotes an electron present on the left (right) dot. 
The dots are coupled by a tunneling parameter $T$, which provides the primary transport channel within our system, while the phonon bath provides a secondary channel of transport provided that the tunneling amplitude is non-zero. 

After applying a Jordan-Wigner transformation \cite{Schaller14} and restricting to the Coulomb blockade regime, the Hamiltonian governing the full system is given by (we take $\hbar=1$ throughout) 
\begin{equation}\label{fullH}
H=H_{\rm S} + H_{\rm E_{ph}} + H_{\rm E_{\rm L}}+ H_{\rm E_{\rm R}} + H_{\rm I_{ph}}+H_{\rm I_{\rm L}}+H_{\rm I_{\rm R}},
\end{equation} 
where
\begin{equation}\label{HS}
H_{\rm S}=\epsilon_{\rm L} \ket{L}\bra{L}+\epsilon_{\rm R} \ket{R}\bra{R} + T\left(\ket{L}\bra{R}+ \ket{R}\bra{L}\right)
\end{equation}
represents the DQD system. The Hamiltonian governing the environments has been split into components $H_{\rm E_{ph}}$ representing the bosonic phonon bath contributions and $H_{{\rm E}_{j}}$ ($j={\rm L,R}$) representing the fermionic leads:
\begin{align}
H_{\rm E_{ph}}&=\sum_{q} \omega_q a_q^\dagger a_q, \label{HEB}\\
H_{{\rm E}_{j}}&=\sum_{k_j}\epsilon_{k_j} c_{k_j}^\dagger c_{k_j}.\label{HEL}
\end{align}
We consider phonon modes $q$ of frequencies $\omega_q$ and standard bosonic creation (annihilation) operators $a_q^\dagger$ ($a_q$). Similarly, the lead modes $k_j$ 
correspond to energy levels $\epsilon_{k_j}$ and standard fermionic creation (annihilation) operators $c_{k_j}^\dagger$ ($c_{k_j}$). Finally, the interaction Hamiltonians are given explicitly as (again $j={\rm L,R}$)
\begin{align}
H_{\rm I_{ph}}&=\sum_q g_q \left(\ket{L}\bra{L}-\ket{R}\bra{R}\right)(a^\dagger_q + a_q),\label{HIB}\\
H_{{\rm I}_{j}}&=\sum_{k_j}\ket{0}\bra{j}t_{k_j}c_{k_j}^\dagger + \ket{j}\bra{0}t_{k_j}^* c_{k_j}.\label{HIL}
\end{align}
Here, each phonon mode $q$ is assumed to couple with strength $g_q$ to the dot occupation difference 
\cite{Brandes07,Stones17}. 
The tunnel coupling strengths between each dot $j$ and the associated lead modes $k_j$ are labelled $t_{k_j}$. 

\subsection{Weak-coupling master equation}\label{wcme}

We begin by presenting a standard BM master equation for the DQD, whereby all environments are considered to be weakly interacting with the system. This will form one aspect of our comparisons later on, in order to highlight the impact of the strong phonon coupling physics captured within the RC treatment. We derive each environmental contribution individually in this case, as they are additive within the BM approximations. Beginning with the lead contributions, the nanojunction has two tunneling transport channels 
defined through the 
energy eigenstates of our system Hamiltonian, $\ket{\Upsilon_1}$ and $\ket{\Upsilon_2}$, which have energies $\Upsilon_1$ and $\Upsilon_2$,  respectively:
\begin{align}
\ket{\Upsilon_1}&=\frac{1}{\sqrt{T^2+\alpha^2}}~\big(\alpha\ket{R}-T\ket{L}\big)\label{E1},\\
\ket{\Upsilon_2}&=\frac{1}{\sqrt{T^2+\alpha^2}}~\big(\alpha\ket{L}+T\ket{R}\big),\label{E2}\\
\Upsilon_1&=\frac{\epsilon_{\rm L}+\epsilon_{\rm R}}{2}+\sqrt{\Delta^2+ T^2},\\
\Upsilon_2&=\frac{\epsilon_{\rm L}+\epsilon_{\rm R}}{2}-\sqrt{\Delta^2+ T^2}. 
\end{align}
Here, $\Delta=({\epsilon_{\rm L}-\epsilon_{\rm R}})/{2}$ and $\alpha=\Delta-\sqrt{\Delta^2+T^2}$. In order to derive a standard BM master equation we 
express our site basis states $\ket{L}$ and $\ket{R}$ in terms of these eigenstates. The left lead interaction Hamiltonian can be written as
\begin{align}
H_{\rm I_{\rm L}}=A_1 \otimes B_1 + A_2 \otimes B_2,
\end{align}
where  $A_1=\alpha\ket{0}\bra{\Upsilon_2}-T\ket{0}\bra{\Upsilon_1}$, $A_2=\alpha\ket{\Upsilon_2}\bra{0}-T\ket{\Upsilon_1}\bra{0}$,  
$B_1=\sum_{k_ {\rm L}}t_{k_ {\rm L}} c^\dagger_{k_{\rm L}}$ and $B_2=\sum_{k_ {\rm L}}t^*_{k_ {\rm L}} c_{k_{\rm L}}$. 
Moving into the interaction picture with respect to $H_{\rm S}+H_{\rm E_{\rm L}}$ we 
trace out the lead degrees of freedom within the standard BM approximations. On returning to the Schr\"odinger picture we find the left lead weak-coupling Liouvillian 
\begin{align}\label{LeftLeadCont}
\mathcal{L}_{\mathrm{L}}\rho_{\rm S}(t)&=-\int^\infty_0 d\tau [A_1,A_2(-\tau)\rho_{\rm S}(t)]C_{12}(\tau)\nonumber\\&+[\rho_{\rm S}(t)A_1(-\tau),A_2]C_{12}(-\tau)\nonumber\\&+[A_2,A_1(-\tau)\rho_{\rm S}(t)]C_{21}(\tau)\nonumber\\&+ [\rho_{\rm S}(t)A_2(-\tau),A_1]C_{21}(-\tau),
\end{align}
where $\rho_{\rm S}(t)$ is the reduced density operator of the DQD system. Here, interaction picture system operators and lead correlation functions are given as
\begin{align}
A_1(-\tau)&=\alpha\ket{0}\bra{\Upsilon_2}e^{i\Upsilon_2 \tau}-T\ket{0}\bra{\Upsilon_1}e^{ i\Upsilon_1 \tau},\nonumber\\
A_2(-\tau)&=\alpha\ket{\Upsilon_2}\bra{0}e^{-i\Upsilon_2 \tau}-T\ket{\Upsilon_1}\bra{0}e^{-i\Upsilon_1 \tau},\nonumber\\
C_{12}(\tau)&=\sum_{k_{\rm L}}|t_{k_{\rm L}}|^2 e^{i\epsilon_{k_{\rm L}}\tau} f_{\rm L}(\epsilon_{k_{\rm L}}),\nonumber\\
C_{21}(\tau)&=\sum_{k_{\rm L}}|t_{k_{\rm L}}|^2 e^{-i\epsilon_{k_{\rm L}}\tau} (1-f_{\rm L}(\epsilon_{k_{\rm L}})),\nonumber
\end{align}
with $f_j(\epsilon)=(e^{\beta_{\rm el}(\epsilon-\mu_j)}+1)^{-1}$ for lead $j$ with chemical potential $\mu_j$ and inverse temperature $\beta_{\rm el}$. 
Further details of this procedure are given in Appendix~\ref{AppA}, as is the explicit form of $\mathcal{L}_{\mathrm{L}}\rho_{\rm S}(t)$ in the site basis.

For the right lead contributions we have an analogous result 
\begin{align}\label{RightLeadCont}
\mathcal{L}_{\mathrm{R}}\rho_{\rm S}(t)&=-\int^\infty_0 d\tau [A_3,A_4(-\tau)\rho_{\rm S}(t)]C_{34}(\tau)\nonumber\\&+[\rho_{\rm S}(t)A_3(-\tau),A_4]C_{34}(-\tau)\nonumber\\&+[A_4,A_3(-\tau)\rho_{\rm S}(t)]C_{43}(\tau)\nonumber\\&+ [\rho_{\rm S}(t)A_4(-\tau),A_3]C_{43}(-\tau),
\end{align}
with
\begin{align}
A_3(-\tau)&=\alpha\ket{0}\bra{\Upsilon_1}e^{i\Upsilon_1 \tau}+T\ket{0}\bra{\Upsilon_2}e^{ i\Upsilon_2 \tau},\nonumber\\
A_4(-\tau)&=\alpha\ket{\Upsilon_1}\bra{0}e^{-i\Upsilon_1 \tau}+T\ket{\Upsilon_2}\bra{0}e^{-i\Upsilon_2 \tau},\nonumber\\
C_{34}(\tau)&=\sum_{k_{\rm R}}|t_{k_{\rm R}}|^2 e^{i\epsilon_{k_{\rm R}}\tau} f_{\rm R}(\epsilon_{k_{\rm R}}),\nonumber\\
C_{43}(\tau)&=\sum_{k_{\rm R}}|t_{k_{\rm R}}|^2 e^{-i\epsilon_{k_{\rm R}}\tau} (1-f_{\rm R}(\epsilon_{k_{\rm R}})).
\end{align}
Again, the explicit form of $\mathcal{L}_{\mathrm{R}}\rho_{\rm S}(t)$ in the site basis is given in Appendix~\ref{AppA}.


The phonon contributions are more easily evaluated directly in terms of the site basis states, where 
\begin{align}
H_{\rm I_{\rm ph}}=A_{5}\otimes B_{5},
\end{align}
with $A_{5}=\ket{L}\bra{L}-\ket{R}\bra{R}$ and $B_{5}=\sum_{q}g_q (a^\dagger_q +a_q)$. Moving into the interaction picture with respect to $H_{\rm S}+H_{\rm E_{ph}}$ and making use of the Baker-Campbell-Hausdorff formula \cite{Sattinger13}, we find 
\begin{align}
A_{5}(t)&=\frac{1}{\zeta ^2}\Big[(\Delta^2+ T^2 {\rm cos}(2\zeta t))\sigma_z
+\Delta T (1-{\rm cos}(2\zeta t))\sigma_x 
\nonumber\\ 
&+\zeta T {\rm sin}(2 \zeta t)\sigma_y 
\Big],\nonumber
\end{align}
where $\zeta=\sqrt{\Delta^2 + T^2}$ and $\sigma_i$ are the usual Pauli matrices in the system subspace $\{\ket{L},\ket{R}\}$, with $\sigma_z=\ket{L}\bra{L}-\ket{R}\bra{R}$. 
Following the standard BM procedure once more, we derive the weak-coupling phonon Liouvillian 
\begin{align}\label{rhoph}
\mathcal{L}_{\mathrm{ph}}\rho_{\rm S}(t)=-\Big\{& M_z [\sigma_z,[\sigma_z,\rho_{\rm S}(t)]]+ M_y [\{\sigma_z,\rho_{\rm S}(t)\},\sigma_y]\nonumber \\&+ M_x [\sigma_z,[\sigma_x,\rho_{\rm S}(t)]]\Big\},
\end{align}
where $M_z=\frac{2\Delta^2 \pi \lambda}{\zeta^2 \beta_{\rm ph} \omega_0}+ \frac{2 \pi T^2}{\zeta^2}J(2\zeta)\mathrm{coth}(\beta_{\rm ph} \zeta)$, $M_y=\frac{2 i \pi T J(2\zeta)}{\zeta}$ and $M_x=\frac{2\Delta T \pi \lambda}{\zeta^2 \beta_{\rm ph} \omega_0}-\frac{2\pi \Delta T}{\zeta^2}J(2\zeta)\mathrm{coth}(\beta_{\rm ph} E)$, for inverse phonon temperature $\beta_{\rm ph}$. 
Throughout this work we consider a Drude-Lorentz spectral density for the phonon bath, defined as 
\begin{equation}
J(\omega)=\sum_q |g_q|^2 \delta(\omega-\omega_q)=\frac{1}{\pi}\frac{2 \lambda \omega \omega_0^2 \gamma}{(\omega_0^2 - \omega^2)+\gamma^2 \omega^2}.
\end{equation}
Here $\lambda$ is the reorganisation energy of the phonons, $\omega_0$ is the central peak 
and $\gamma$ is a parameter which tunes the spectral density from sharply peaked to broad. 
The lead spectral densities are defined analogously as 
\begin{equation}
{\mathcal J}_{\rm L/R} (\omega)=\sum_{k_{\rm{L/R}}}|t_{k_{\rm L/R}}|^2\delta (\omega-\epsilon_{\rm{L}/R}),
\end{equation}
with both flat and Lorentzian forms considered below. 
Combining Eqs.~(\ref{LeftLeadCont}), (\ref{RightLeadCont}) and (\ref{rhoph}) we arrive at the weak-coupling BM master equation 
\begin{align}\label{WCME}
\dot\rho_{\rm S}(t)&=-i[H_{\rm S},\rho_{\rm S}(t)]+\mathcal{L}_{\mathrm{L}}\rho_{\rm S}(t)+\mathcal{L}_{\mathrm{R}}\rho_{\rm S}(t)+\mathcal{L}_{\mathrm{ph}}\rho_{\rm S}(t), 
\end{align}
which on imposing the infinite bias limit becomes equivalent (for our spectral density) to the weak-coupling approaches considered for example in Refs.~\cite{Stones17,Brandes07}.
Our concern is primarily with finite bias, however.

\subsection{Implementing the RC mapping}\label{RCMap}

\begin{figure}
\center
\includegraphics[width=\linewidth]{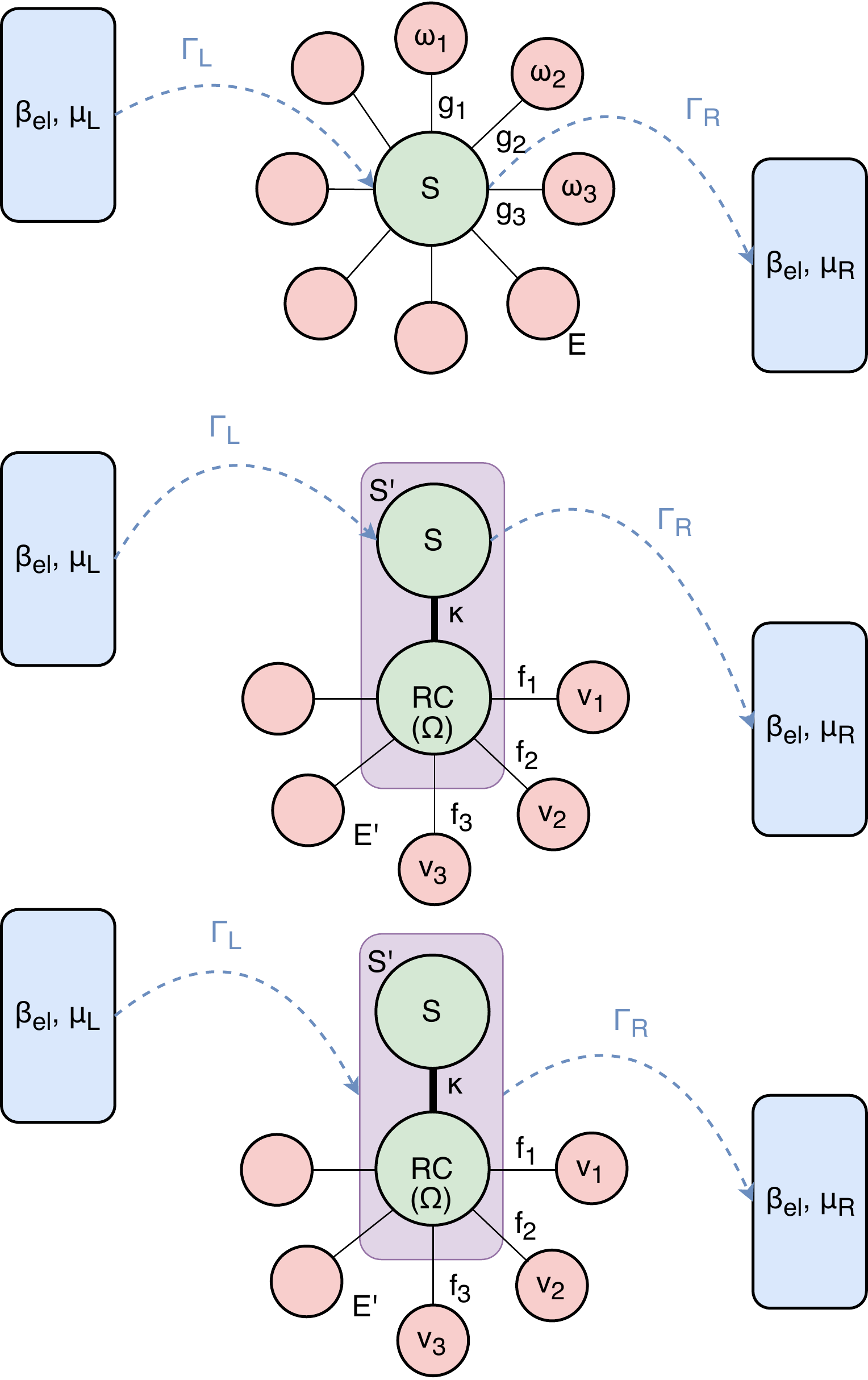}
\caption{{Schematic diagram of the weak-coupling and RC approaches. 
Top: The weak-coupling approach,  where S and E represent the system and phonon environment, respectively, in the original framework, with couplings $g_q$ and environmental frequencies $\omega_q$. Middle: The additive treatment of the RC mapping, with S$'$ depicting the augmented system obtained after the mapping. This is comprised of the original system S and the RC (of frequency $\Omega$) interacting through $\kappa$, while E$'$ represents the residual environment with frequencies $\nu_k$ and couplings $f_k$. Note that the leads couple to the original system in the additive case. Bottom: The non-additive treatment of the RC mapping in which the leads couple to the augmented system S$'$. Other symbols are as defined in Fig.~\ref{diagram1}.} 
}
\label{RCDiagram}
\end{figure}

Having looked at the weakly-coupled master equation in the previous section we now consider the RC formalism and how it can be used to treat {the same model though with strong electron-phonon coupling. To implement the RC mapping we apply a unitary normal mode transformation to our Hamiltonian, $H$, that incorporates a collective environmental degree of freedom (the RC) into the system (see Fig.~\ref{RCDiagram}), as previously applied to phonon coupling in other contexts~\cite{IlesSmith14,IlesSmith16,Strasberg16,Garg85,Newman17, Maguire18}.} The strong coupling between our original system, S, and the RC results in a repartitioning of the original system and environment. As such we attain an augmented system, S$'$, which includes both the original system and the RC, along with an explicit interaction between the two (denoted by $\kappa$). We treat the residual environment, E$'$, as being weakly coupled to the RC and incorporate its influence using standard BM methods.

{The RC approach has been successfully benchmarked against the numerically-exact hierarchical equations of motion over a range of parameters, including both broad and peaked spectral densities \cite{IlesSmith14, IlesSmith16} and situations in which system-environment correlations and non-Markovian effects are pronounced. For low temperatures, the residual environment correlation time may become extended such that the BM approximation on E$'$ no longer applies. We do not consider such temperature regimes here, however.}

{On applying the RC mapping as described to $H$ of Eq.~(\ref{fullH}) we find 
\begin{align}
H&=H_{\rm S}+H_{\rm E_{ph}}+H_{\rm I_{ph}}+H_{\rm E_{\rm L}}+ H_{\rm E_{\rm R}} +H_{\rm I_{\rm L}}+H_{\rm I_{\rm R}}
\nonumber\\
&\rightarrow H_{\rm S'}+H_{\rm E'}+H_{\rm I'}+H_{\rm E_{\rm L}}+ H_{\rm E_{\rm R}} +H_{\rm I_{\rm L}}+H_{\rm I_{\rm R}}.
\end{align} 
Here, the augmented system Hamiltonian becomes
\begin{align}\label{HSR}
H_{\rm S'}&= \epsilon_{\rm L}\ket{L}\bra{L}+\epsilon_{\rm R}\ket{R}\Bra{R}+ T(\ket{L}\bra{R}+\ket{R}\bra{L}) \nonumber\\&+ \kappa(\ket{L}\bra{L}-\ket{R}\bra{R}) (a^\dagger + a) + \Omega a^\dagger a,
\end{align}
where the first line contains terms associated with the original system, and we now also have its coupling to the RC as well as the self energy of the latter in the second line. The RC (collective coordinate) is defined directly from the electron-phonon interaction term in Eq.~(\ref{HIB}) such that
\begin{equation}
\kappa(a^\dagger + a)= \sum_q g_q (a_q^\dagger + a_q),
\end{equation}
where the RC creation ($a^{\dagger}$) and annihilation ($a$) operators satisfy $[a,a^{\dagger}]=1$, $\kappa^2=\sum_q g_q^2$ and the RC frequency is $\Omega=\omega_0$~\cite{IlesSmith14,IlesSmith16}. 
The residual phonon environment E$'$ is defined in terms of bosonic creation and annihilation operators $b_k^\dagger$ and $b_k$, respectively, for modes of frequency $\nu_k$,
\begin{equation}\label{Her}
H_{{\rm E}'}=\sum_k \nu_k b_k^\dagger b_k, 
\end{equation}
and is coupled with strengths $f_k$ directly to the RC 
by{~\cite{IlesSmith14,IlesSmith16,Hartmann00}} 
\begin{align}\label{Hir}
H_{{\rm I}'}&= (a^\dagger + a)\sum_k f_k (b_k^\dagger+ b_k) + (a^\dagger + a)^2 \sum_k \frac{f_k^2}{\nu_k}.
\end{align}
The RC mapping has no impact on the lead operators and so Eqs.~(\ref{HEL}) and~(\ref{HIL}) remain unchanged. We thus denote the total mapped Hamiltonian by
\begin{equation}
H'=H_{\rm S'}+H_{\rm E'}+H_{\rm I'}+H_{\rm E_L}+H_{\rm E_R}+H_{\rm I_L}+H_{\rm I_R}.
\end{equation}}

Having outlined the mapping at the Hamiltonian level, we now consider the dynamical treatment of phonon interactions within the RC framework, before moving on to examine the lead couplings to our augmented system. The RC master equation has previously been derived in several different contexts~\cite{IlesSmith14,IlesSmith16,Strasberg16,Newman17} and gives a Liouvillian of the form
\begin{align}\label{phonme}
\mathcal{L}_{\rm RC}\rho_{\rm S'}(t)&=-i[H_{\rm S'},\rho_{\rm S'}(t)]-[A_{\mathrm{ph}},[\chi_{\mathrm{ph}},\rho_{\rm S'}(t)]]\nonumber\\&+[A_{\mathrm{ph}},\{\phi_{\mathrm{ph}},\rho_{\rm S'}(t)\}].
\end{align}
Here 
$\rho_{\rm S'}(t)$ is the reduced density operator of the augmented system (comprising both S and RC), 
$A_{\rm ph}=(a^\dagger + a)$, and
\begin{align}
\chi_{\mathrm{ph}}&=\int^\infty_0\int^\infty_0 d\omega d\tau J_{\rm RC}(\omega) \mathrm{cos}(\omega \tau)\mathrm{coth}\Big(\frac{\beta_{\rm ph} \omega}{2}\Big) A_{\mathrm{ph}}(-\tau) \nonumber\\
&\approx \frac{\pi}{2} \sum_{j k} J_{\rm RC}(\eta_{jk})\mathrm{coth}\Big(\frac{\beta_{\rm ph} \eta_{jk}}{2}\Big) A_{\mathrm{ph}_{jk}}\ket{\psi_j}\bra{\psi_k},\\
\phi_{\mathrm{ph}}&=\int^\infty_0\int^\infty_0 d\omega d\tau \frac{J_{\rm RC}(\omega) \mathrm{cos}(\omega \tau)}{\omega} [H_{\rm S'},A_{\mathrm{ph}}(-\tau)]\nonumber\\
&\approx \frac{\pi}{2} \sum_{j k} J_{\rm RC}(\eta_{jk}) A_{\mathrm{ph}_{jk}}\ket{\psi_j}\bra{\psi_k},
\end{align}
with $H_{\rm S'}|\psi_j\rangle=\psi_j|\psi_j\rangle$, $A_{\mathrm{ph}_{jk}}=\langle\psi_j|A_{\rm ph}|\psi_k\rangle$, $\eta_{jk}=\psi_j-\psi_k$ and $J_{\rm RC}=(1/2\pi)\omega e^{-\omega/\Lambda}$. 
{In practice, we represent the augmented system 
within a basis consisting of Fock states for the RC mode and the double quantum dot states $\{|0\rangle,|L\rangle,|R\rangle\}$. We take fourteen Fock states in all our calculations, 
which ensures 
numerical convergence.} 
For the sake of completeness, we include a derivation of the RC Liouvillian within Appendix~\ref{AppB}. 

We now turn our attention to the lead couplings and derive their contributions when coupled to our augmented system. 
It is common to simplify the model by taking the infinite (source-drain) lead bias limit, $\mu_{\rm L}=-\mu_{\rm R}\rightarrow\infty$, see e.g.~Refs.~\cite{Stones17,Brandes07,Harbola06}. 
In this regime, provided the lead spectral densities are flat, it is accurate to treat them 
by simply adding the standard Liouvillian terms to the master equation [see Eqs.~(\ref{LIb}) and~(\ref{RIb})] unmodified due to the (potentially strong) phonon interactions. 
However, we wish to extend our considerations 
to finite bias regimes, in which case we shall see that including non-additive contributions becomes crucial. In order to do this we must derive the lead couplings with respect to our augmented system. 
In the following we examine the left lead contributions explicitly; the treatment of the right lead is analogous and so shall be stated afterwards. 
Setting $A'_1=\ket{0}\bra{L}$ and $A'_2 =\ket{L}\bra{0}$ 
allows us to write the left lead interaction Hamiltonian, Eq.~(\ref{HIL}), in the form
\begin{equation}\label{leftleadHi}
H_{\rm I'}^{\mathrm{Left}}(t)= A'_1(t)\otimes B_1(t) + A'_2(t)\otimes B_2(t).
\end{equation}
Here we have moved into the interaction picture with respect to the \emph{mapped} system Hamiltonian $H_{\rm S'}$, i.e. $A'_{1(2)}(t)=e^{iH_{\rm S'}t}A_{1(2)}'e^{-iH_{\rm S'}t}$, with $B_{1(2)}(t)=e^{iH_{\rm E_L}t}B_{1(2)}e^{-iH_{\rm E_L}t}$ as before. Now, following the standard BM approach~\cite{Schaller14} within the mapped representation, 
we arrive at a Schr\"odinger picture Liouvillian of the form 
\begin{align}\label{ferme2}
{\mathcal L}'_{\rm L}\rho_{\rm S'}&=-\sum_{k_{\rm L}}|t_{k_{\rm L}}|^2\int^\infty_0 d\tau [A'_1,A'_2(-\tau)\rho_{\rm S'}]e^{i\epsilon_{k_{\rm L}}\tau} f_{\rm L}(\epsilon_{k_{\rm L}})\nonumber\\&+[\rho_{\rm S'}A'_2(-\tau),A'_1]e^{i\epsilon_{k_{\rm L}}\tau} (1 - f_{\rm L}(\epsilon_{k_{\rm L}}))\nonumber\\
&+[A'_2,A'_1(-\tau) \rho_{\rm S'}]e^{-i\epsilon_{k_{\rm L}}\tau} (1- f_{\rm L}(\epsilon_{k_{\rm L}}))\nonumber\\&+[\rho_{\rm S'} A'_1(-\tau),A'_2]e^{-i\epsilon_{k_{\rm L}}\tau}  f_{\rm L}(\epsilon_{k_{\rm L}}),
\end{align}
where $\rho_{\rm S'}(t)$ is again the reduced density operator of the mapped (augmented) system comprising of both the DQD and the RC. 
As with the phonon case, we express the system operators 
within the eigenbasis of $H_{\rm S'}$:
\begin{align}\label{leftleadteigen}
A'_1(-\tau)&=\sum_{jk}A'_{1_{j,k}}e^{-i\eta_{jk}\tau}\ket{\psi_j}\bra{\psi_k},\\
A'_2(-\tau)&=\sum_{jk}A'_{2_{j,k}}e^{-i\eta_{jk}\tau}\ket{\psi_j}\bra{\psi_k}.
\end{align}
Here, $\eta_{jk}$ and $\ket{\psi_j}$ are defined as before, 
with $A'_{1_{j,k}}=\bra{\psi_j}A'_1\ket{\psi_k}$ and $A'_{2_{j,k}}=\bra{\psi_j}A'_2\ket{\psi_k}$. We also consider the continuum limit for the lead coupling, taking $\Gamma_{\rm L}(\omega)= 2 \pi\sum_{k_{\rm L}}|t_{k_{\rm L}}|^2\delta(\omega-\epsilon_{k_{\rm L}})=2\pi J_{\mathrm{L}}(\omega)$ to give
\begin{align}
{\mathcal L}'_{\rm L}\rho_{\rm S'}&=- \sum_{jk}\Bigg[\frac{\Gamma_{\rm L}(\eta_{jk})}{2}\Big([A'_1,f_{\rm L}(\eta_{jk})A'_{2_{j,k}}\ket{\psi_j}\bra{\psi_k} \rho_{\rm S'}]\nonumber\\&+[\rho_{\rm S'} A'_{2_{j,k}}\ket{\psi_j}\bra{\psi_k}(1-f_{\rm L}(\eta_{jk})),A'_1]\Big)\nonumber\\&+ \frac{\Gamma_{\rm L}(-\eta_{jk})}{2}\Big([A'_2,(1-f_{\rm L}(-\eta_{jk}))A'_{1_{j,k}}\ket{\psi_j}\bra{\psi_k} \rho_{\rm S'}]\nonumber\\&+[\rho_{\rm S'} A'_{1_{j,k}}\ket{\psi_j}\bra{\psi_k}f_{\rm L}(-\eta_{jk}),A'_2]\Big)\Bigg].
\end{align}
To simplify this expression we now consider a flat lead spectral density, such that $\Gamma_{\rm L} (-\eta_{jk})=\Gamma_{\rm L} (\eta_{jk})=\Gamma_{\rm L}$. We also define
\begin{align}
\chi_1&=\sum_{jk} A'_{1_{j,k}}\ket{\psi_j}\bra{\psi_k}f_{\rm L}(-\eta_{jk})\\ \phi_1&= \sum_{jk} A'_{1_{j,k}}\ket{\psi_j}\bra{\psi_k}(1 - f_{\rm L}(-\eta_{jk}))\\
\chi_2&=\sum_{jk}  A'_{2_{j,k}}\ket{\psi_j}\bra{\psi_k}f_{\rm L}(\eta_{jk})\\ \phi_2&=\sum_{jk}  A'_{2_{j,k}}\ket{\psi_j}\bra{\psi_k}(1 - f_{\rm L}(\eta_{jk})),
\end{align}
which allows the Liouvillian to be written in the more compact form
\begin{align}\label{leftme}
\mathcal{L}'_{\mathrm{L}}\rho_{\rm S'}(t)&=-\Gamma_{\rm L} \Big([A'_1,\chi_2 \rho_{\rm S'}(t)]+[\rho_{\rm S'}(t)\phi_2,A'_1]\nonumber\\&+[A'_2,\phi_1 \rho_{\rm S'}(t)]+[\rho_{\rm S'}(t)\chi_1,A'_2]\Big).
\end{align}
The right lead Liouvillian can be given analogously as
\begin{align}\label{rightme}
\mathcal{L}'_{\mathrm{R}}\rho_{\rm S'}(t)&=-\Gamma_{\rm R} \Big([A'_3,\chi_4 \rho_{\rm S'}(t)]+[\rho_{\rm S'}(t)\phi_4,A'_3]\nonumber\\&+[A'_4,\phi_3 \rho_{\rm S'}(t)]+[\rho_{\rm S'}(t)\chi_3,A'_4]\Big),
\end{align}
where $A'_3=\ket{0}\bra{R}$, $A'_4=\ket{R}\bra{0}$, and we define 
\begin{align}
\chi_3&=\sum_{jk} A'_{3_{j,k}}\ket{\psi_j}\bra{\psi_k}f_{\rm R}(-\eta_{jk})\\ \phi_3&= \sum_{jk} A'_{3_{j,k}}\ket{\psi_j}\bra{\psi_k}(1 - f_{\rm R}(-\eta_{jk}))\\
\chi_4&=\sum_{jk}  A'_{4_{j,k}}\ket{\psi_j}\bra{\psi_k}f_{\rm R}(\eta_{jk})\\ \phi_4&=\sum_{jk}  A'_{4_{j,k}}\ket{\psi_j}\bra{\psi_k}(1 - f_{\rm R}(\eta_{jk})),
\end{align}
with $A'_{3_{j,k}}=\bra{\psi_j}A'_3\ket{\psi_k}$ and $A'_{4_{j,k}}=\bra{\psi_j}A'_4\ket{\psi_k}$.
Finally we combine Eqs.~(\ref{phonme}), 
(\ref{leftme}) and (\ref{rightme}) to arrive at the full master equation
\begin{align}\label{fullme}
\dot\rho_{\rm S'}(t)&=-i[H_{\rm S'},\rho_{\rm S'}(t)]\nonumber\\&-\Gamma_{\rm L} \Big([A'_1,\chi_2 \rho_{\rm S'}(t)]+[\rho_{\rm S'}(t)\phi_2,A'_1]\nonumber\\&+[A'_2,\phi_1 \rho_{\rm S'}(t)]+[\rho_{\rm S'}(t)\chi_1,A'_2]\Big)\nonumber\\&-\Gamma_{\rm R} \Big([A'_3,\chi_4 \rho_{\rm S'}(t)]+[\rho_{\rm S'}(t)\phi_4,A'_3]\nonumber\\&+[A'_4,\phi_3 \rho_{\rm S'}(t)]+[\rho_{\rm S'}(t)\chi_3,A'_4]\Big)\nonumber\\&-[A_{\rm ph},[\chi_{\rm ph},\rho_{\rm S'}(t)]]+[A_{\rm ph},\{\phi_{\rm ph},\rho_{\rm S'}(t)\}].
\end{align}
Though this is additive at the level of the augmented system S$'$ due to the BM approximations made for both the lead couplings and the residual phonon bath, it encodes non-additive contributions at the level of the original system $S$ due to the explicit presence of coupling to the RC and the resultant changes in the lead Liouvillians. 
On the other hand, a truly additive master equation is obtained following the RC mapping if we simply append the standard weak-coupling lead Liouvillians, Eqs.~(\ref{LeftLeadCont}) and~(\ref{RightLeadCont}), to the phonon Liouvillian of Eq.~(\ref{phonme}): 
\begin{align}\label{addme}
\dot\rho_{\rm S'}(t)&=-i[H_{\rm S'},\rho_{\rm S'}(t)]\nonumber\\&+(\mathcal{L}_{\mathrm{L}}\otimes\openone_{\rm RC})\rho_{\rm S'}(t)+(\mathcal{L}_{\mathrm{R}}\otimes\openone_{\rm RC})\rho_{\rm S'}(t)\nonumber\\&-[A_{\rm ph},[\chi_{\rm ph},\rho_{\rm S'}(t)]]+[A_{\rm ph},\{\phi_{\rm ph},\rho_{\rm S'}(t)\}],
\end{align}
where $\openone_{\rm RC}$ is the identity operator in the RC subspace. 
This procedure implements the RC mapping for the phonons, but the lead dissipators are now unaware of the augmented system and so are treated phenomenologically. We shall see that this results in inaccuracies for flat leads when we consider finite bias regimes, and even at infinite bias for Lorentzian lead spectral densities. 

\subsection{Full counting statistics}\label{FCS}


Electron counting statistics provides a means to track the electron flow between our system and a reservoir of our choosing. 
In this work we consider 
the transfer of electrons between the right lead and the right dot, 
though we note that within the steady-state it would be equivalent to track the flow between the left lead and the left dot. 
As mentioned previously, the fact that the RC treatment results in a Markovian master equation in the mapped representation for our originally non-Markovian problem implies that the counting statistics formalism can be applied in the standard manner.
We thus consider single electron transitions at each stage of the evolution, and as such we can take the description of a generic reduced density operator $\dot{\rho}(t)=\mathcal{L}\rho(t)$  with appropriate Liouvillian superoperator $\mathcal{L}$, and split it into three separate terms
\begin{equation}\label{CSex}
\dot{\rho}^{(n)}(t)=\mathcal{L}_0\rho^{(n)}(t) + \mathcal{I}^+ \rho^{(n-1)}(t)+ \mathcal{I}^-\rho^{(n+1)}(t).
\end{equation} 
Here we explicitly consider the number of electrons present on the right lead during each transition, as denoted by the superscripts present on $\rho(t)$. The first term includes the cases in which no electron transfers between the right dot and the right lead occur, thus the number of electrons present on the lead, $n$, is unaffected. The remaining two terms consider the cases of electron movement onto and off the right lead, such that $\mathcal{I}^+$ is the rate at which the right lead gains electrons ($n-1 \rightarrow n$) and $\mathcal{I}^-$ is the rate at which it loses them ($n+1 \rightarrow n$).

While it is possible to compute the full probability distribution  $P_n(t)= \mathrm{Tr}(\rho^{(n)}(t))$  of the electron tunneling events, it is generally computationally expensive and time consuming. It is often more useful to consider the cumulant generating function $\mathcal{C}(\Psi,t)$, which is defined as
\begin{equation}
e^{\mathcal{C}(\Psi,t)}=\sum_{n=-\infty}^{n=\infty}P_n(t) e^{in\Psi}.
\end{equation}
We can then calculate the cumulants of our charge distribution using $\mathcal{C}(\Psi,t)$. We focus specifically on the first and second order cumulants as these are proportional to 
the mean particle current and zero-frequency noise 
when taking time derivatives: 
{
\begin{align}\label{current&noise}
\mathrm{Current}&=e \langle\langle I^1\rangle\rangle= e \frac{d}{dt}\Big(\frac{\partial\mathcal{C}(\Psi,t)}{\partial(i\Psi)}\Big)\bigg\rvert_{\Psi=0,t\rightarrow\infty}\\
\mathrm{Noise}&=2e^2\langle\langle I^2\rangle\rangle= 2e^2\frac{d}{dt}\Big(\frac{\partial^2\mathcal{C}(\Psi,t)}{\partial(i\Psi)^2}\Big)\bigg\rvert_{\Psi=0,t\rightarrow\infty}.
\end{align}}

Following a rigorous derivation discussed in \cite{Flindt10} we are able to define current and noise in an alternative way:
\begin{align}\label{current}
\langle\langle I^1\rangle\rangle&=\langle\langle\tilde{0}|\mathcal{I}|0\rangle\rangle,\\\label{noise}
\langle\langle I^2\rangle\rangle&=\langle\langle\tilde{0}| \mathcal{J}|0\rangle\rangle - 2\langle\langle\tilde{0}| \mathcal{IRI}|0\rangle\rangle,
\end{align}
where $\mathcal{I}=\mathcal{I}^+-\mathcal{I}^-$, $\mathcal{J}=\mathcal{I}^++\mathcal{I}^-$, $|0\rangle\rangle $ is the right eigenvector of our Liouvillian corresponding to the zero eigenvalue, and $\langle\langle\tilde{0}|$ is the identity operator (the left eigenvector of the zero eigenvalue). We have also introduced the pseudoinverse of our Liouvillian, $\mathcal{R}=\mathcal{QL}^{-1}\mathcal{Q}$, with projectors $\mathcal{P}= |0\rangle\rangle\langle\langle\tilde{0}| $ and $\mathcal{Q}=\openone-\mathcal{P}$. The pseudoinverse is well defined as the inversion is performed in the subspace spanned by $\mathcal{Q}$. For more details see \cite{Flindt05,Flindt10}.

We shall also examine the Fano factor of our system, $F$, 
defined as the ratio between the noise and the current, i.e.~a comparison between the variance of the data and its mean:
\begin{equation}
F=\frac{\langle\langle I^2\rangle\rangle}{\langle\langle I^1\rangle\rangle}.
\end{equation}
The Fano factor can be used to signal deviations of our current fluctuations from Poissonian statistics. For $F> 1$ we have super-Poissonian behaviour, indicating that the transport channel is disordered. This means that there will be times whereby few electrons tunnel into the right lead, and then at other times there will be a rapid succession of electron tunneling events. For $F< 1$ we have sub-Poissonian behaviour. In this case the tunneling of electrons between the dots and the right lead is very uniform, and as such electrons appear to tunnel at evenly spaced intervals. As a final point we note the changes to Eqs.~(\ref{current}) and (\ref{noise}) when we are in the infinite bias regime, in which case $\mathcal{I}=\mathcal{J}=\mathcal{I}^+$ so we have
\begin{align}
\langle\langle I^1\rangle\rangle&=\langle\langle\tilde{0}|\mathcal{I}^+|0\rangle\rangle,\\
\langle\langle I^2\rangle\rangle&=\langle\langle\tilde{0}| \mathcal{I}^+|0\rangle\rangle - 2\langle\langle\tilde{0}| \mathcal{I^+RI^+}|0\rangle\rangle.
\end{align}

\section{Results and Discussion}\label{Results}

\begin{figure*}[t]
\center
     \subfloat[ ][Infinite Bias]{{\includegraphics[width=0.45\textwidth]{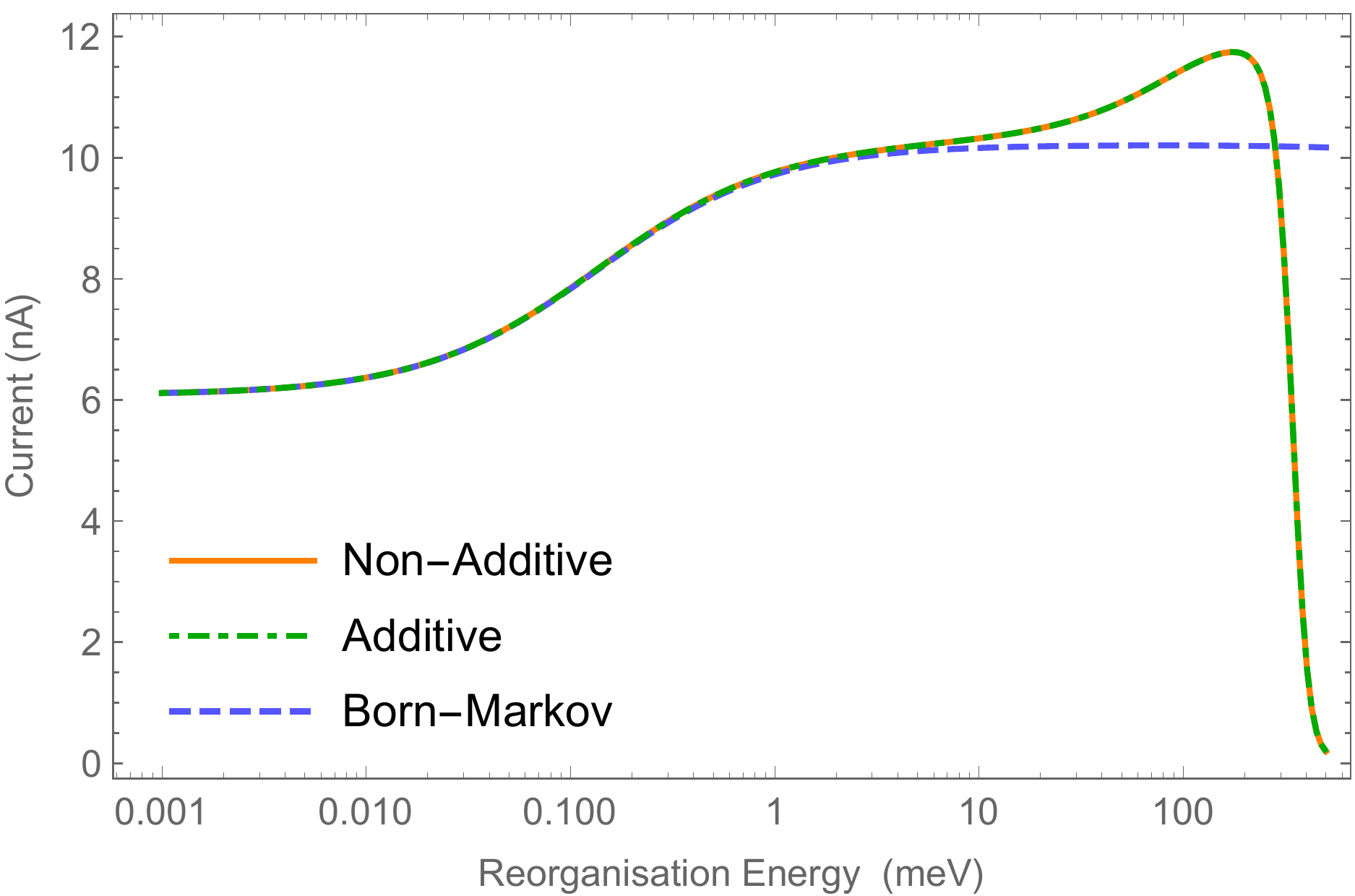}}}%
     \qquad    
     \subfloat[ ][Finite Bias]{{\includegraphics[width=0.45\textwidth]{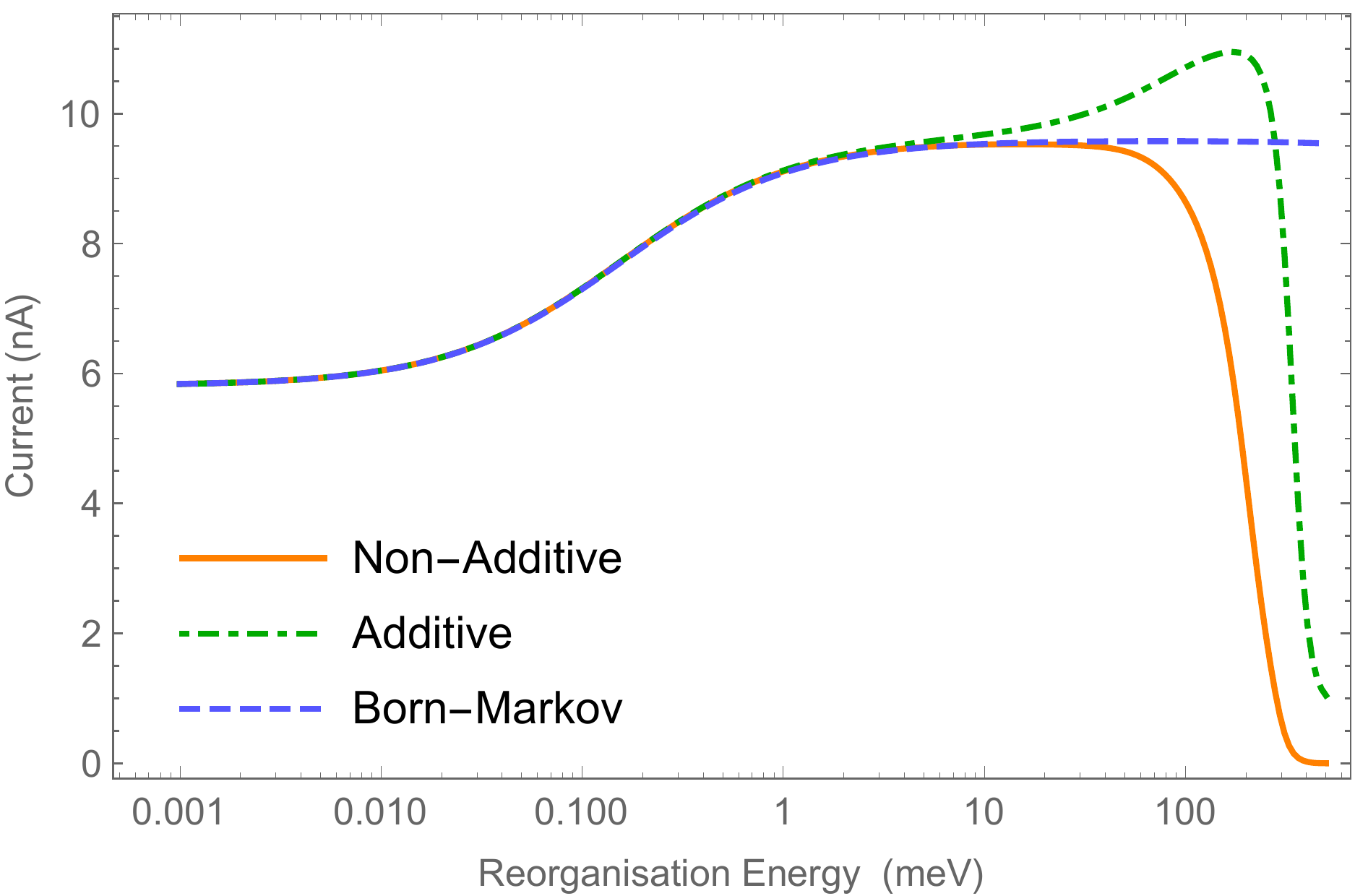}}}%
     \qquad   \\
    \subfloat[ ][Low Bias]{{\includegraphics[width=0.45\textwidth]{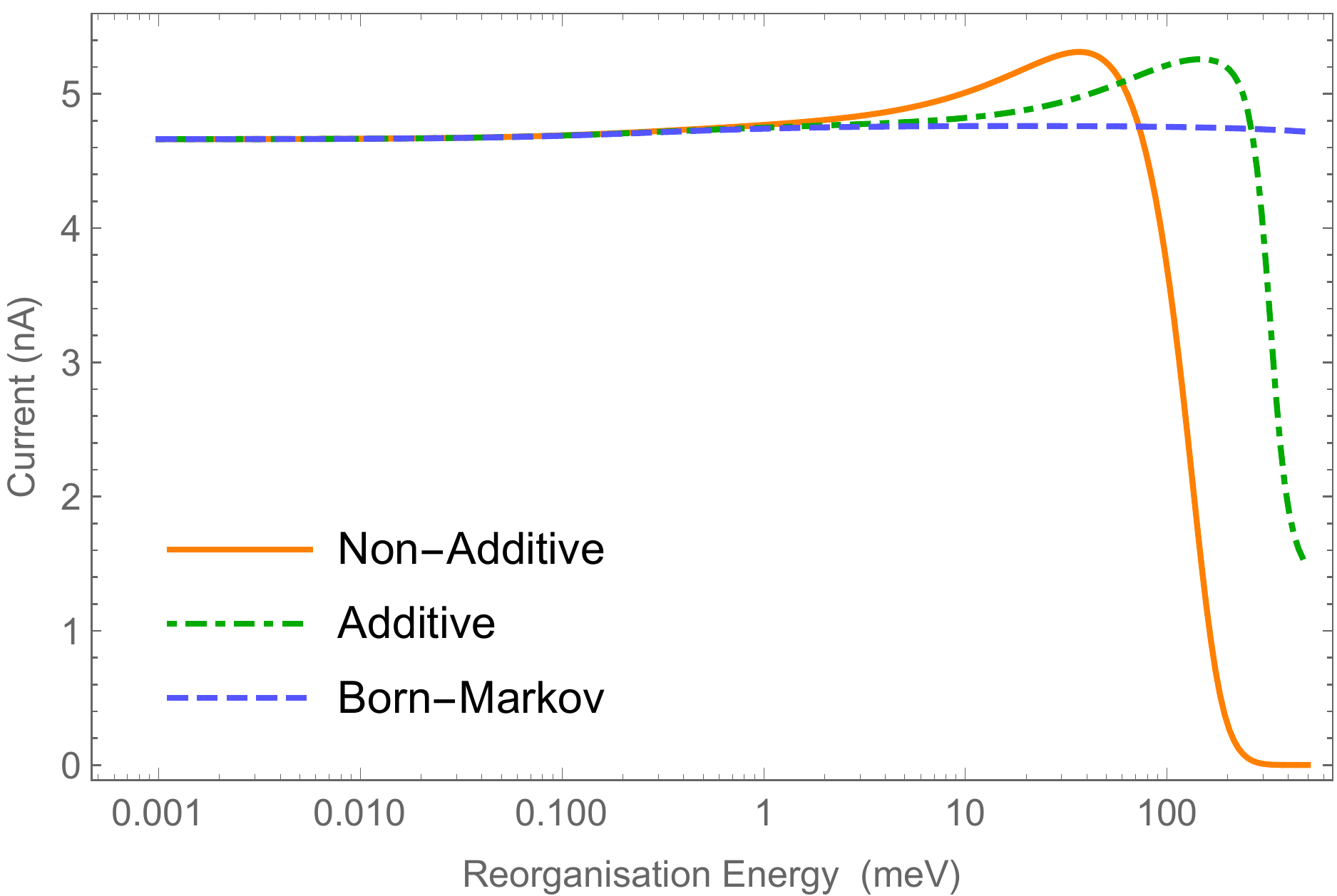}}}%
 \qquad    
     \subfloat[ ][Zero Bias]{{\includegraphics[width=0.45\textwidth]{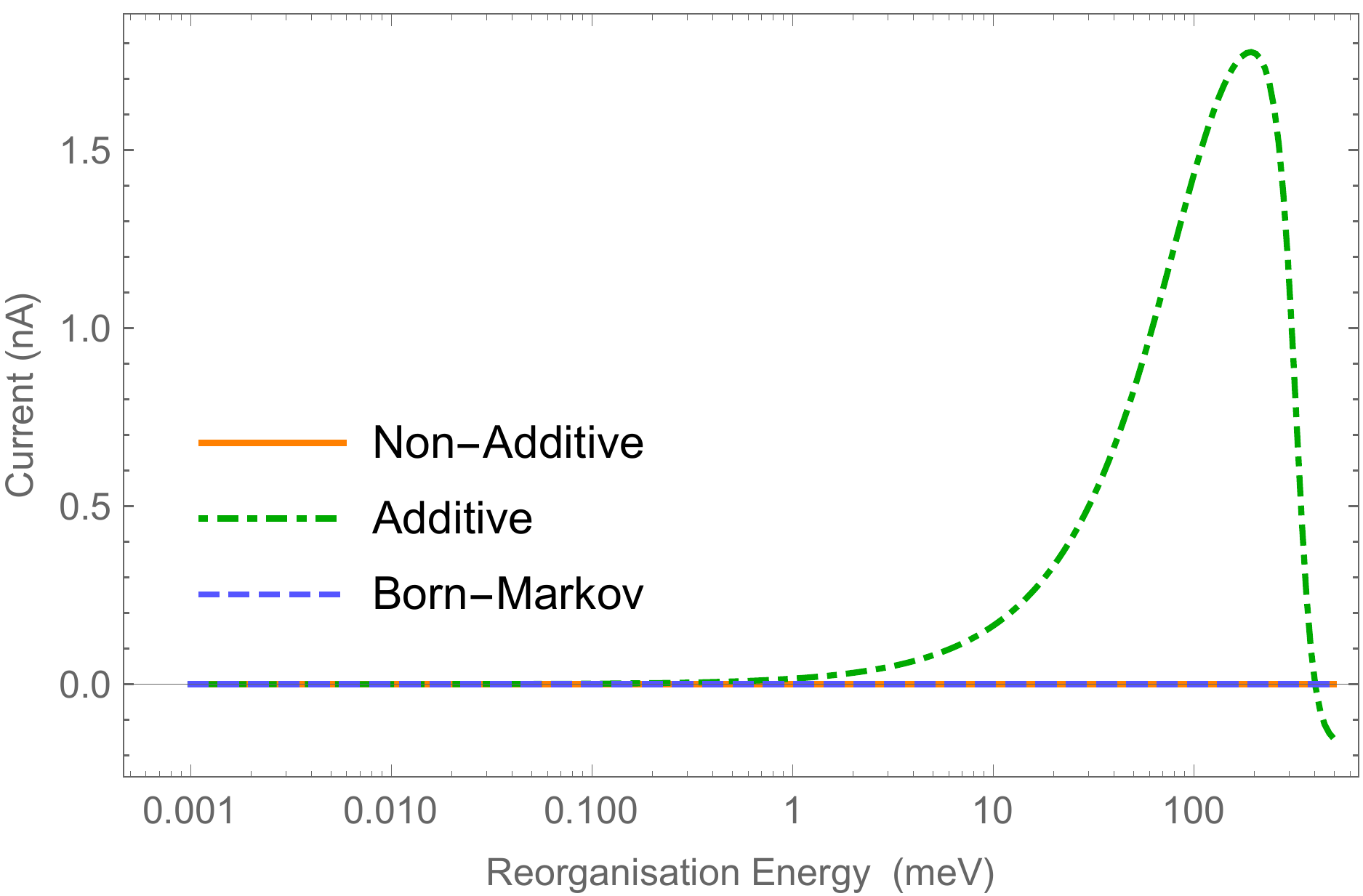}}}%
     \caption{\label{Currentvslambda}Current flow as a function of reorganisation energy $\lambda$ calculated within (a) the infinite bias regime (taken as $\mu_{\rm L}=-\mu_{\rm R}=10,000$~meV), (b) a finite bias regime ($\mu_{\rm L}=-\mu_{\rm R}=250$~{meV}), (c) a low bias regime  ($\mu_{\rm L}=-\mu_{\rm R}=100$~{meV}) where one of the tunneling channels is outside the bias window, and (d) the zero bias case in which ($\mu_{\rm L}=\mu_{\rm R}=100$~{meV}). Other parameters used are $\epsilon_{\rm L}= 150$~{meV}, $\epsilon_{\rm R}=50$~{meV}, $T=100$~{meV}, $1/\beta_{\rm el}=1/\beta_{\rm ph}=300$~{K}, $\omega_0=\gamma=100$~{meV}, and $\Gamma_{\rm L}= \Gamma_{\rm R} =0.1$~{meV}.}%
\end{figure*}

\subsection {Impact of phonon coupling}

\begin{figure*}[t]
     \subfloat[ ][$\lambda=0.1$~meV, $1/\beta_{\rm el}=1/\beta_{\rm ph}=300$~K]{{\includegraphics[width=0.45\textwidth]{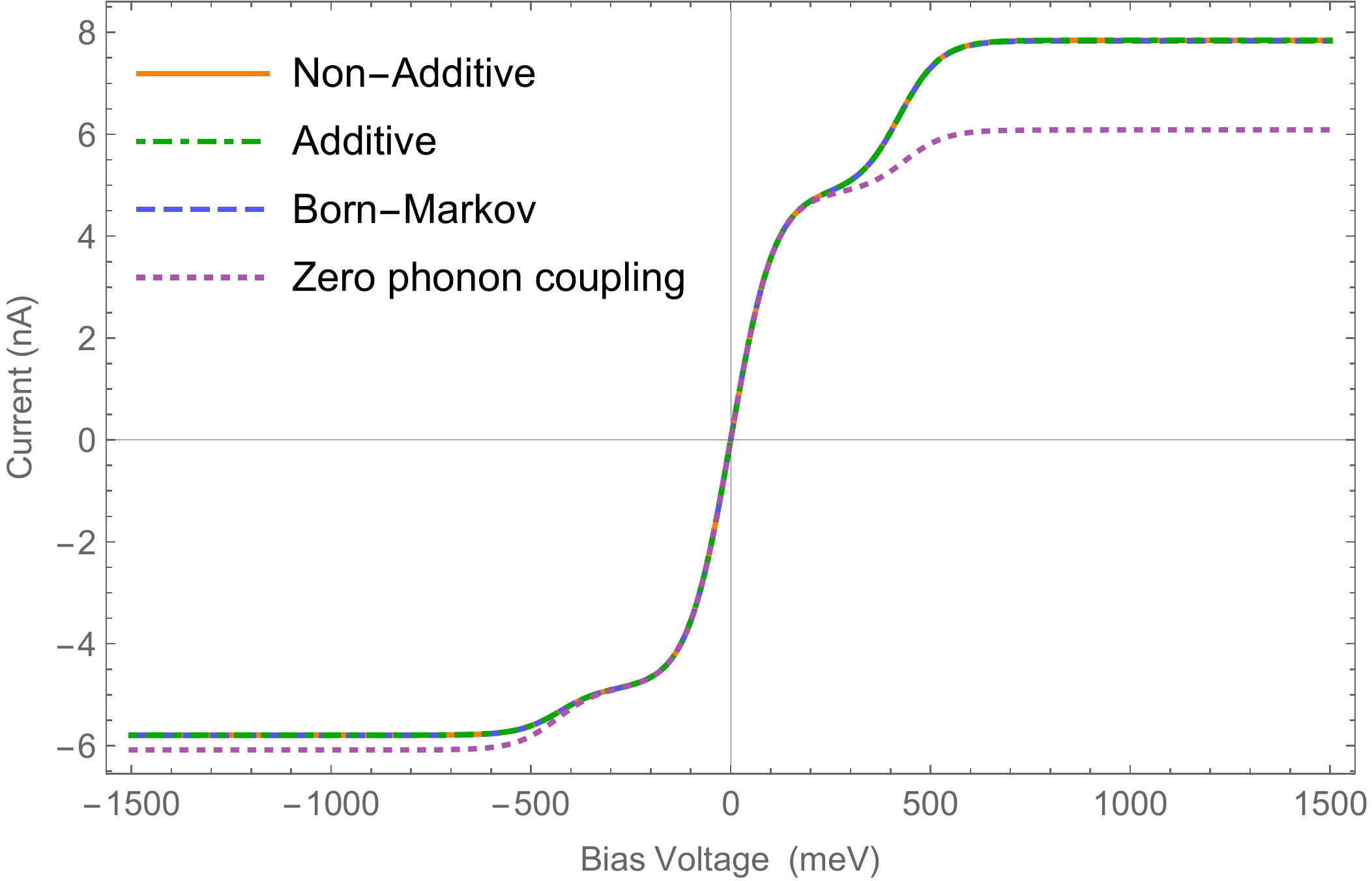}}}%
     \qquad    
     \subfloat[ ][$\lambda=0.1$~meV, $1/\beta_{\rm el}=1/\beta_{\rm ph}=100$~K]{{\includegraphics[width=0.45\textwidth]{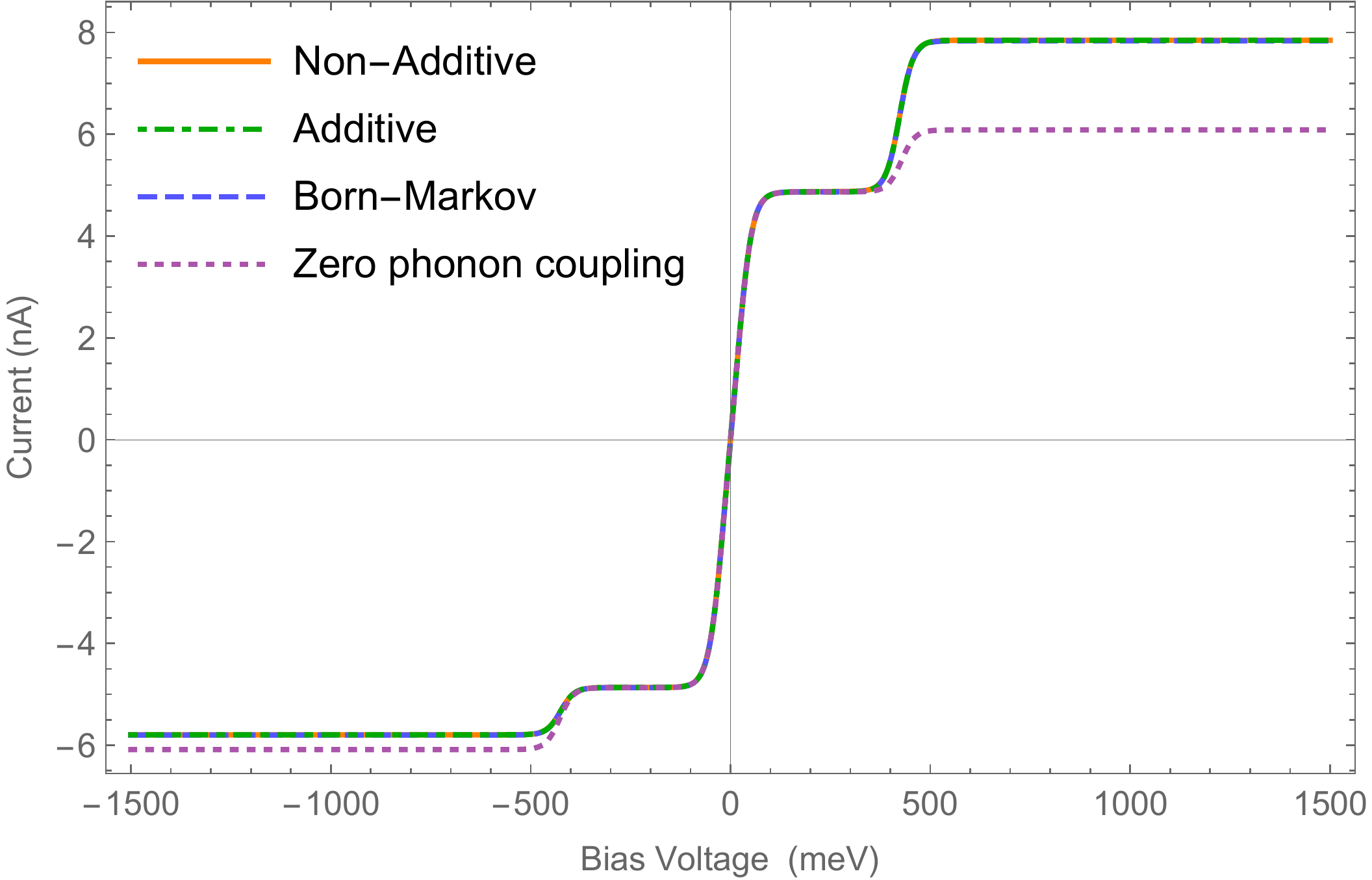}}}%
     \qquad   \\
    \subfloat[ ][$\lambda=200$~meV, $1/\beta_{\rm el}=1/\beta_{\rm ph}=300$~K]{{\includegraphics[width=0.45\textwidth]{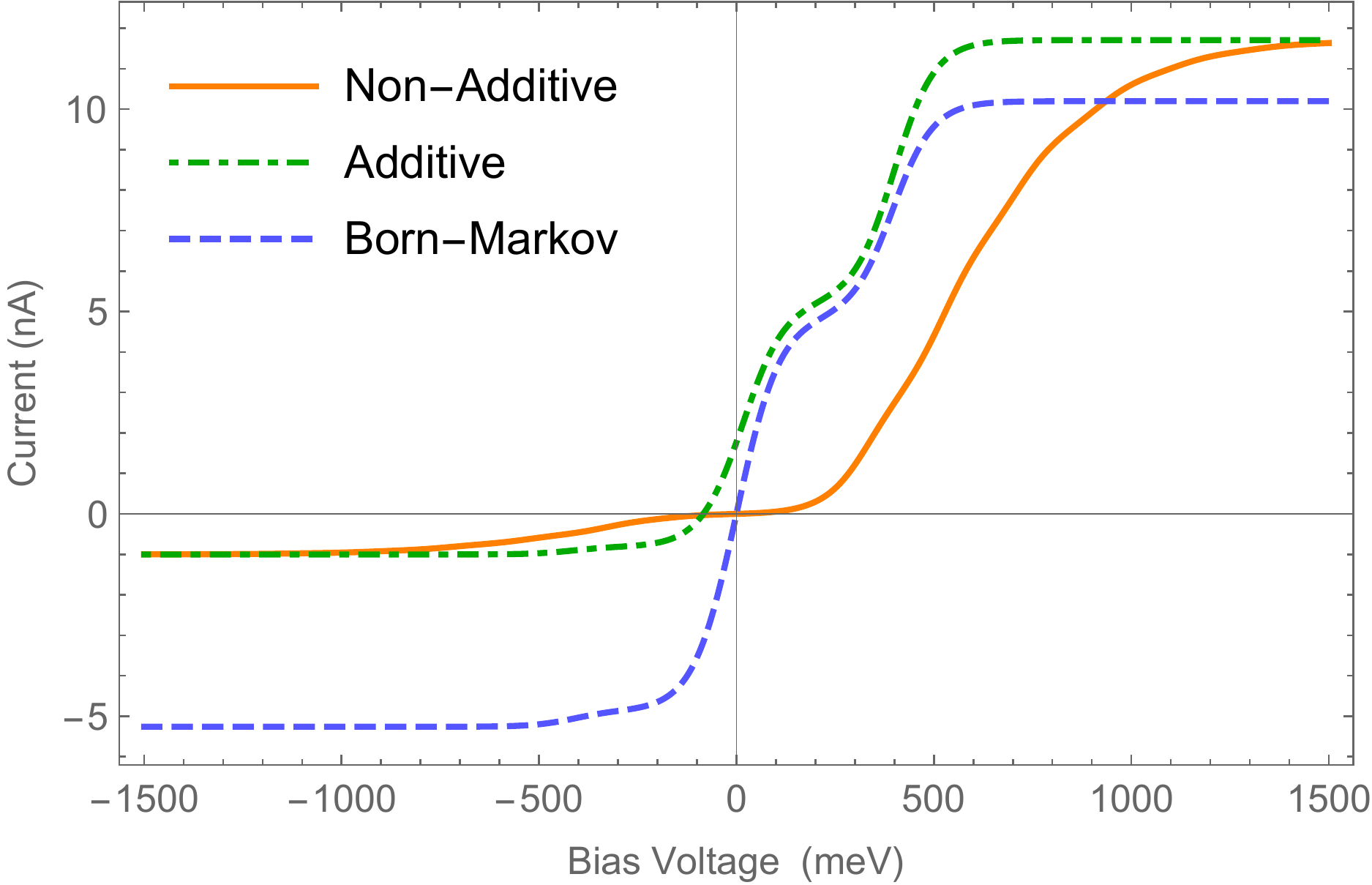}}}%
 \qquad   
     \subfloat[ ][$\lambda=200$~meV, $1/\beta_{\rm el}=1/\beta_{\rm ph}=100$~K]{{\includegraphics[width=0.45\textwidth]{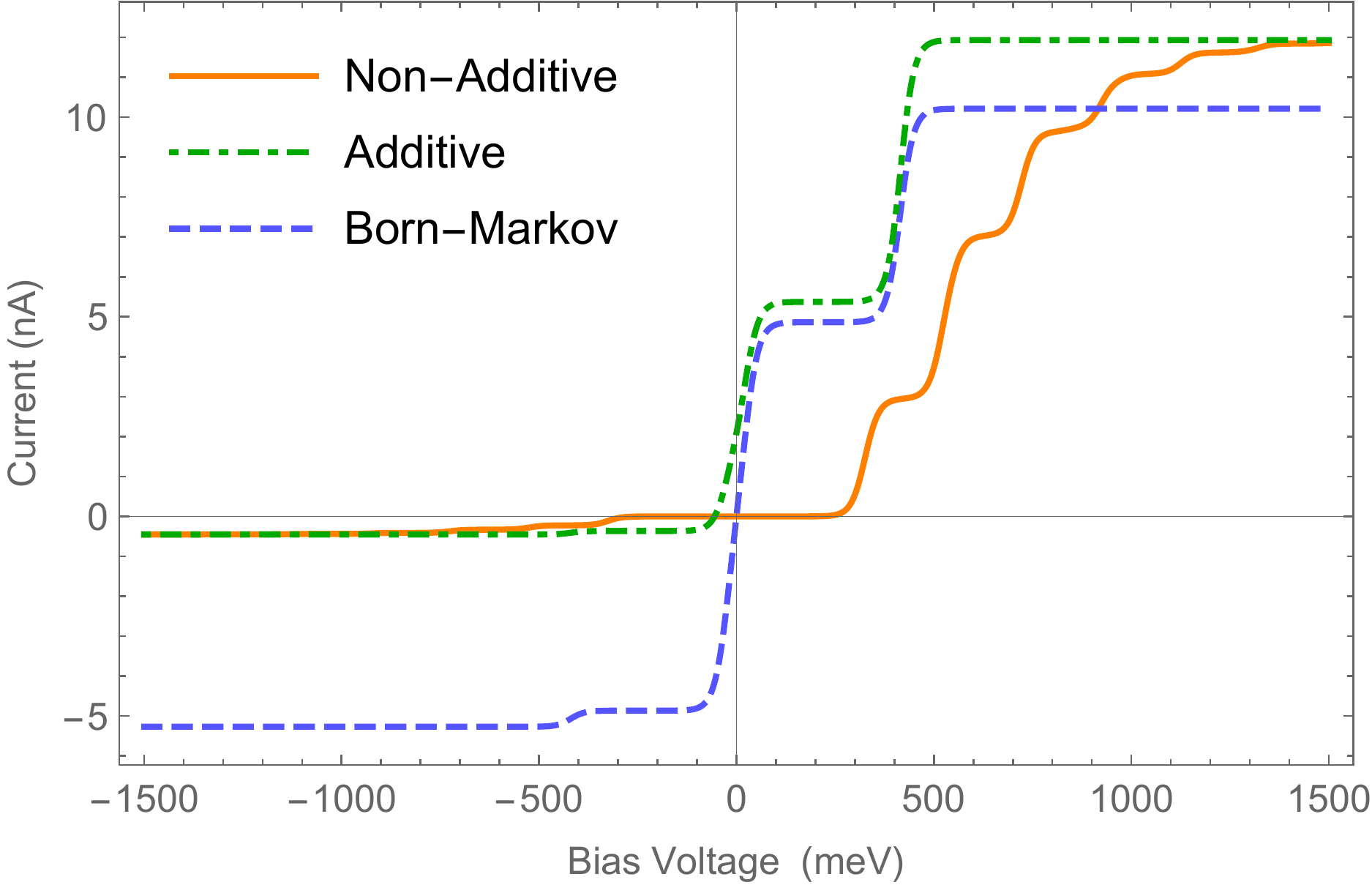}}}%
     \caption{\label{Currentvsbias}Current flow as a function of voltage bias, $eV=\mu_{\rm L}-\mu_{\rm R}$, calculated within (a) a weak coupling regime ($\lambda=0.1$~{meV}) at room temperature ($1/\beta_{\rm el}=1/\beta_{\rm ph}=300$~K), (b) a weak coupling regime ($\lambda=0.1$~{meV}) at a colder temperature ($1/\beta_{\rm el}=1/\beta_{\rm ph}=100$~K), (c) a strong coupling regime ($\lambda=200$~{meV}) at room temperature ($1/\beta_{\rm el}=1/\beta_{\rm ph}=300$~K) and (d) a strong coupling regime ($\lambda=200$~{meV}) at a colder temperature ($1/\beta_{\rm el}=1/\beta_{\rm ph}=100$~K). Other parameters used are $\epsilon_{\rm L}= 150$~{meV}, $\epsilon_{\rm R}=50$~{meV}, $T=100$~{meV}, $\omega_0=\gamma=100$~{meV}, and $\Gamma_{\rm L}= \Gamma_{\rm R} =0.1$~{meV}. 
}%
\end{figure*}

We begin our results discussion by considering the impact that the phonon coupling strength has on the overall current flowing into the right lead, when both leads are taken to be flat, ${\mathcal J}_{\rm L/R}(\omega)=\Gamma_{\rm L/R}/2\pi$. 
We vary the coupling strength by altering the phonon reorganisation energy $\lambda$. In Fig.~\ref{Currentvslambda}(a) we consider the infinite bias limit, whereby we take the chemical potential difference of our leads to be so large that they become an effective source (left lead) and drain (right lead). 
As such we insist on unidirectional flow of electrons, preventing any movement of electrons onto the left lead or off the right lead.
From this figure we see that for the infinite bias regime there are no non-additive effects, and as such the additive (i.e.~phenomenological lead dissipator) RC treatment of Eq.~(\ref{addme}) and the full non-additive master equation [Eq.~(\ref{fullme})] provide complete agreement for any value of $\lambda$. The BM approach displays qualitatively different behaviour at strong reorganisation energies where the weak phonon coupling theory breaks down. As we increase $\lambda$ from zero we see that all three approaches predict a growth in current. In addition to the primary transport mechanism (characterised by the inter-dot tunneling $T$), for larger phonon coupling the system is able to make better use of the secondary phonon-assisted channel, and as such the current flow increases. 
For values of the phonon reorganisation energy that are the same order of magnitude as the dot energy levels we find that the current is enhanced compared to the BM predictions, indicating that only the strong coupling techniques are able to fully utilise both transport channels. 
{For even larger values of $\lambda$ the current flow within the strong coupling theories then drops rapidly. 
In this regime the substantial RC displacements lead to a strong suppression of vibrational overlaps, which 
renormalises the effective inter-dot tunneling strength towards zero (not captured within the BM theory). 
This in turn inhibits electronic transitions and results in localisation of electrons on the left dot, such that no current can flow.}


Moving away from the infinite bias regime we consider cases whereby we relax the stringent tunneling constraints that 
it imposes on our system. In  Fig.~\ref{Currentvslambda}(b) we examine a finite bias value where $\mu_{\rm L}=-\mu_{\rm R}=250$ meV. Here, the left lead chemical potential is still larger than the DQD energies, encouraging transport from left to right, however 
the previous source-drain picture no longer applies. 
Though increasing the phonon coupling from zero again results in an increased current flow, 
the additive treatment 
no longer appears to experience full localisation at very large reorganisation energies. 
In contrast, the non-additive approach again shows complete localisation for very large values of $\lambda$, however, unlike the non-additive case it no longer predicts any improvement in current over the weak-coupling BM master equation. 
It is important to note that within the non-additive treatment, the leads become sensitive to DQD energies that have been shifted by the inclusion of the RC, which is not the case for the additive approach. Here, this shift acts to increase backflow of electrons in comparison to the additive theory, and thus removes the region of improved current flow 
seen at infinite bias, or indeed at lower biases as in Fig.~\ref{Currentvslambda}(c). 

When considering 
Fig.~\ref{Currentvslambda}(c) we note that the 
DQD eigenstate $\ket{\Upsilon_1}$ is now higher in energy than the left lead chemical potential 
and as such it falls outside the bias window. This results in tunnelling through only a single level, 
$\ket{\Upsilon_2}$, 
and thus an overall reduction in current flow. For the BM approach there is now no longer any improvement in current as we increase the phonon coupling, whereas the two strong coupling approaches demonstrate a slight increase for large reorganisation energies, with the non-additive treatment predicting this to occur at lower reorganisation energies than the additive theory. We also see that, as in Fig.~\ref{Currentvslambda}(b), the additive approach does not display the same level of electron localisation as the non-additive theory at very large reorganisation energies, and so does not experience as great a suppression of current.

We attribute the differences in predictions between the non-additive and additive theories at large reorganisation energies to a breakdown of the additive approximation. 
This can be seen explicitly in the case of zero bias, as shown in Fig.~\ref{Currentvslambda}(d). Here, we would expect no current flow, as both leads are held at equal chemical potentials and all three baths (leads and phonons) have the same temperature. As such, there should be no energy gradients for our electrons to follow. This is correctly depicted by the non-additive RC mapping and even the BM approach, though the latter does not more generally apply when the reorganisation energy becomes significant. However, the additive RC case displays markedly incorrect behaviour, predicting a current flow of over $1.5$ nA for large values of the phonon coupling. {This results from the leads being insensitive to 
the strong phonon coupling within the additive treatment, which thus predicts a non-equilibrium steady state even at zero bias.} {Specifically, the true equilibrium state at strong-coupling should be defined with respect to the augmented system of the double dot plus RC. This is accurately captured within the full non-additive treatment, but within the additive approximation the leads couple directly to the dots without reference to the RC. They thus cannot act to equilibrate the system within the correct, augmented basis, which results here in the incorrect generation of a non-equilibrium steady state even in the absence of chemical potential and temperature gradients. The associated spurious current may be positive or negative depending on the energetics of the dots and the phonon coupling, which varies as the reorganisation energy is increased.}

\begin{figure*}[t]
     \subfloat[ ][Infinite Bias]{{\includegraphics[width=0.45\textwidth]{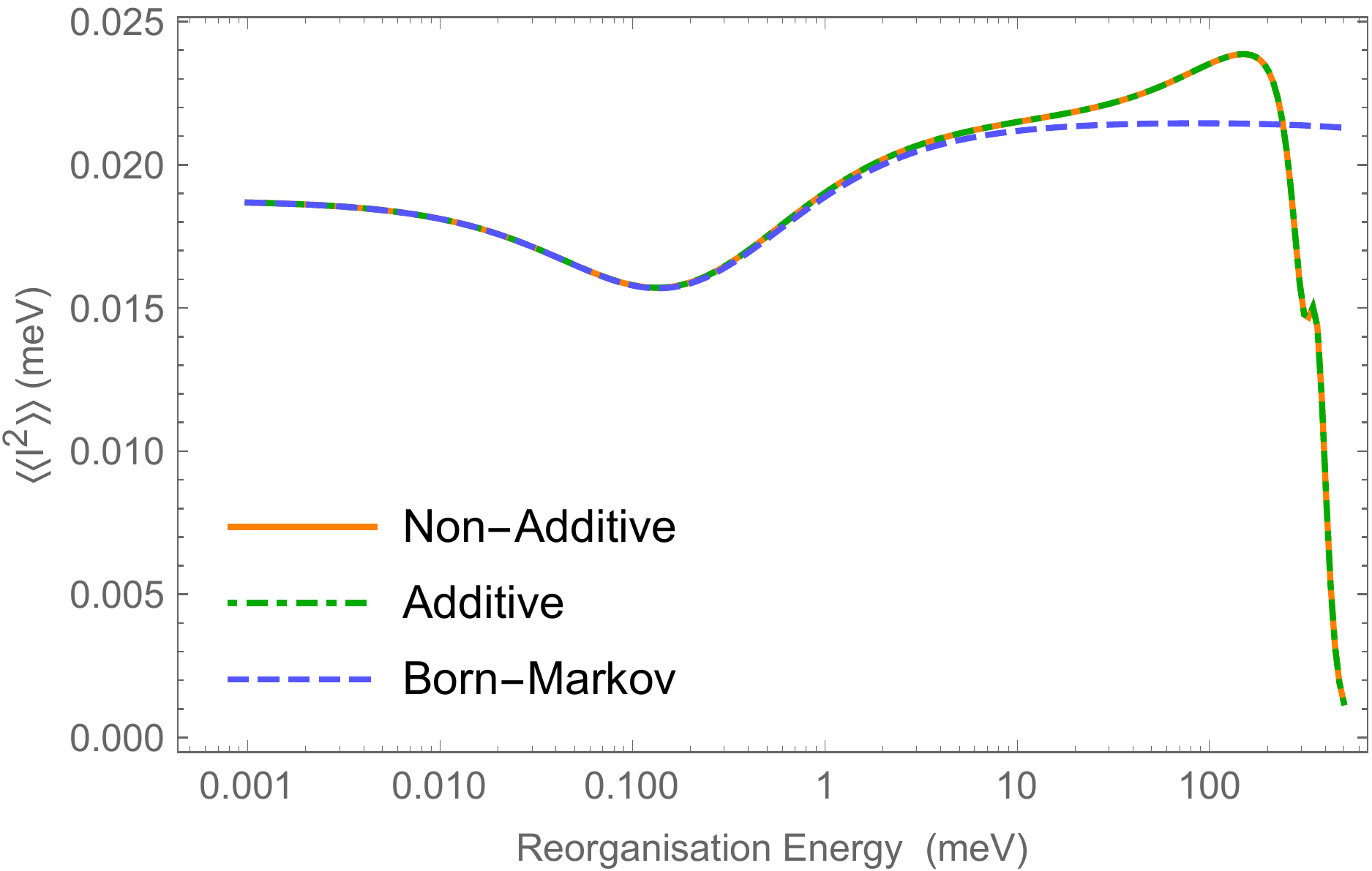}}}%
     \qquad    
     \subfloat[ ][Infinite Bias]{{\includegraphics[width=0.45\textwidth]{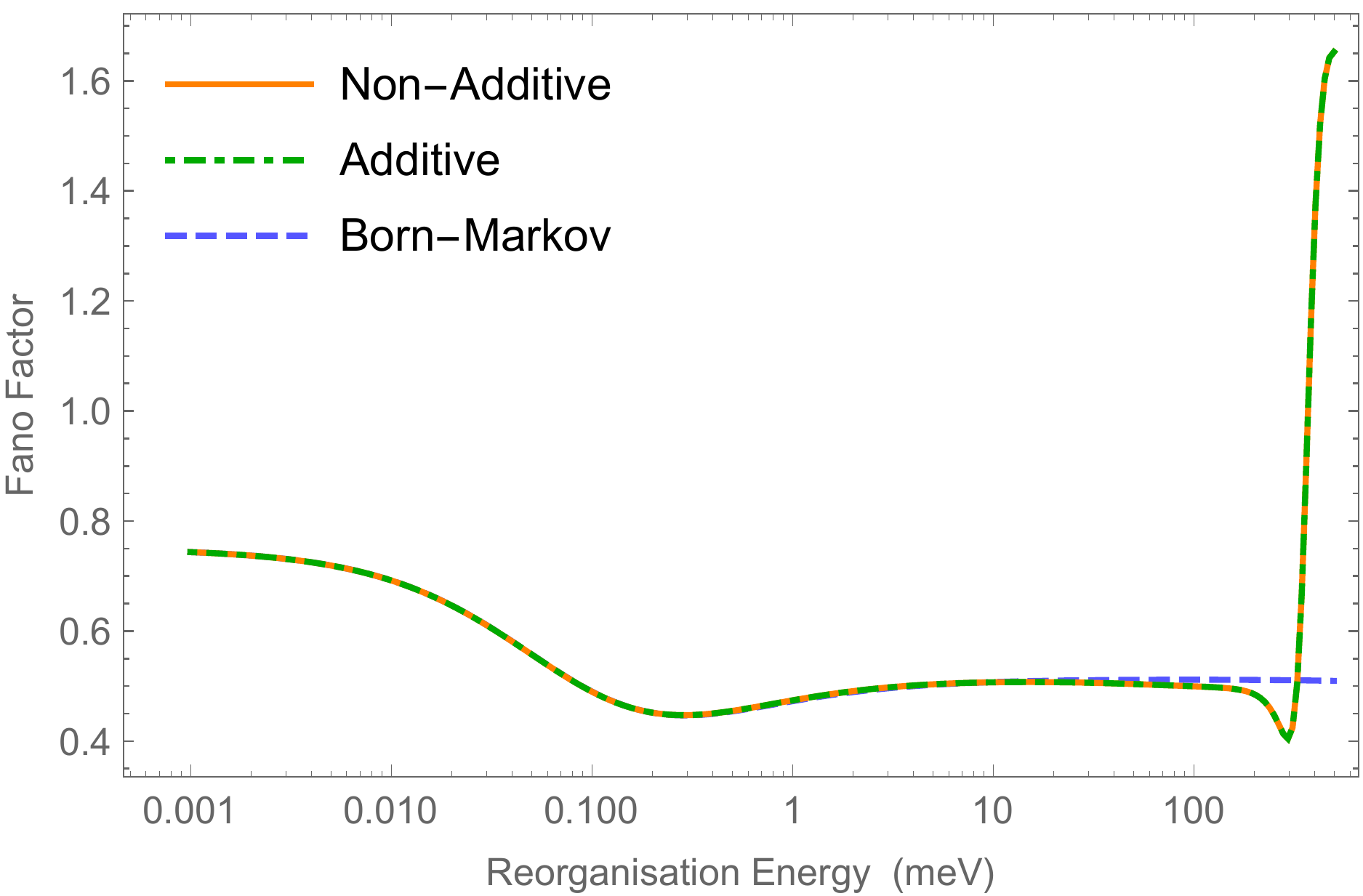}}}%
     \qquad   \\
    \subfloat[ ][Finite Bias]{{\includegraphics[width=0.45\textwidth]{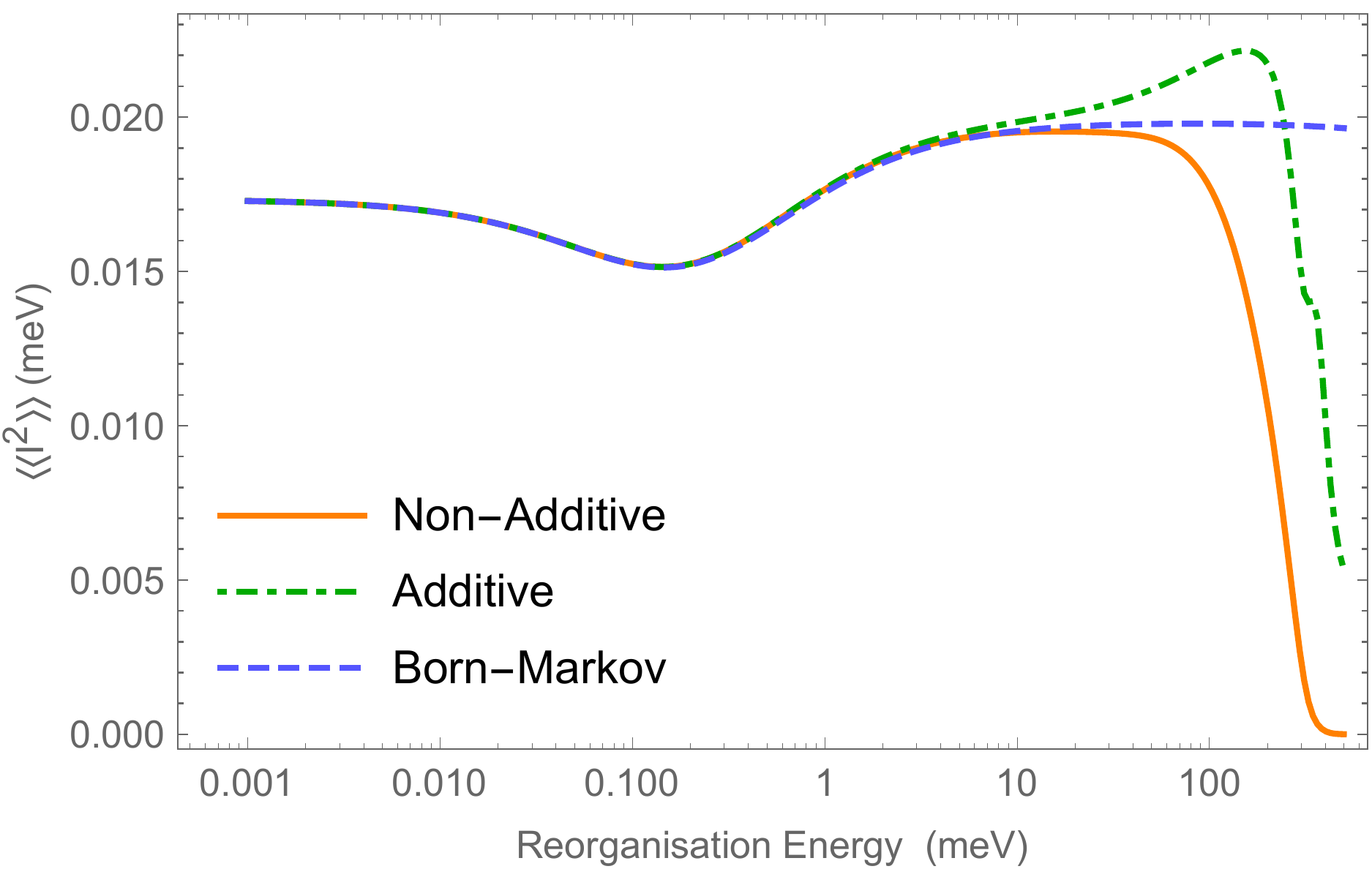}}}%
 \qquad   
     \subfloat[ ][Finite Bias]{{\includegraphics[width=0.45\textwidth]{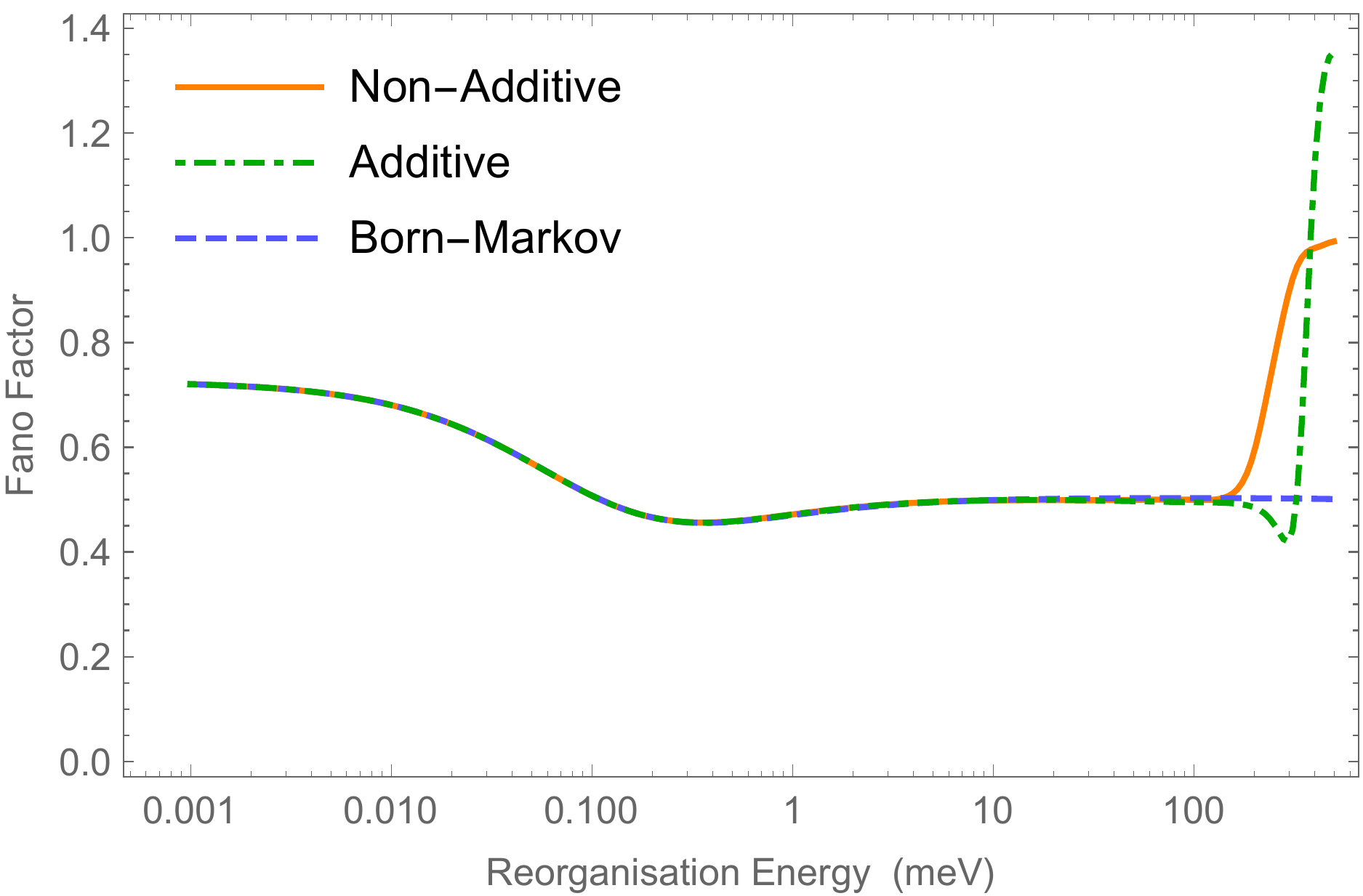}}}%
     \caption{\label{Fanovslambda} {(a) The second cumulant, $\langle\langle I^2\rangle\rangle$, and (b) the Fano factor as a function of phonon reorganisation energy $\lambda$ within the infinite bias limit (taken as $\mu_{\rm L}=-\mu_{\rm R}=10,000$~{meV}). In (c) and (d) a finite bias regime ($\mu_{\rm L}=-\mu_{\rm R}=250$~{meV}) is considered. Other parameters used are $\epsilon_{\rm L}= 150$~{meV}, $\epsilon_{\rm R}=50$~{meV}, $T=100$~{meV}, $1/\beta_{\rm el}=1/\beta_{\rm ph}=300$~{K}, $\omega_0=\gamma=100$~{meV}, and $\Gamma_{\rm L}= \Gamma_{\rm R} =0.1$~{meV}.} 
}%
\end{figure*}

\subsection{Impact of voltage bias}
Having considered the impact that phonon coupling has on the current flow we now turn our attention to the lead chemical potential bias. We have already demonstrated that non-additive contributions can result in significant differences when we consider finite bias regimes, and we aim in this section to consolidate that knowledge and outline further interesting behaviour that can be missed with a phenomenological additive treatment.

We start, however, with Fig.~\ref{Currentvsbias}(a), in which we consider weak phonon coupling 
(with both the leads and the phonon bath held at room temperature) 
and so expect to see complete agreement between all of our approaches. 
Nevertheless, this figure allows us to highlight a number of aspects of the physics of our DQD system. 
Firstly, we note that even when the bias is negative (encouraging right to left transitions) we obtain a lower current flow than in the forward tunneling case. This is a by product of the energy levels present within our DQD system. Although the coupling between each dot and their respective lead is the same, the difference in energies of our dots results in an asymmetry within the system. 

Another key feature of the current-voltage graphs is their usefulness in determining at what bias values our energy levels $\ket{\Upsilon_1}$ and $\ket{\Upsilon_2}$ enter into the bias window. Considering positive bias, from Fig.~\ref{Currentvsbias}(a) we are able to determine two plateau-like regions in the current. 
In the first, we have achieved the maximum current possible for tunneling through the $\ket{\Upsilon_2}$ state. For the parameters considered, this state is held at an energy of roughly $-10$ meV and so is always within the bias window, except around $V= 0$. On the other hand, $\ket{\Upsilon_1}$ has a much larger energy of $220$ meV and is therefore initially inaccessible. When the bias voltage is increased towards $500$ meV we see that the current begins to increase again, as now both system eigenstates lie within the bias window. Further increasing the bias leads to an eventual plateau in the current, as the system is now fully utilising its transport channels. Qualitatively similar behaviour is seen in Fig.~\ref{Currentvsbias}(b) where the lead and phonon temperatures have been reduced, resulting in sharper current features. 

Also shown in Figs.~\ref{Currentvsbias}(a) and (b) are the corresponding current-voltage curves in the absence of phonon coupling.
For positive bias, we see that even a weak phonon coupling strength results in a significant increase in current. Again, this can be attributed to the opening up of the secondary phonon-assisted transport mechanism. For the temperatures considered here phonon emission dominates over absorption, and thus phonons mediate transitions from the higher lying DQD eigenstate $\ket{\Upsilon_1}$ to the lower lying state $\ket{\Upsilon_2}$, resulting in a current enhancement. For the same reason, current is slightly suppressed at negative bias, whereby the phonons preferentially mediate transitions in the opposite direction to the overall flow of current. 

These features are even more pronounced at strong phonon coupling, 
as depicted in Figs.~\ref{Currentvsbias}(c) and (d), where we now also see substantial variations between the different theoretical approaches. Consistent with our previous discussion, the additive approach is inaccurate for small bias values, again giving a finite current even for vanishing bias voltage, though the additive and non-additive theories do tend towards each other in large bias regimes.
Furthermore, we can now see significant discrepancies in predictions for the locations of plateaus within the current curves, particularly between the non-additive technique and the others, as well as a systematic underestimation within the weak phonon coupling approach of the current at large positive bias, and a corresponding overestimation of the current magnitude at large negative bias. The non-additive theory also displays a suppression of current for small bias voltages that is not reproduced by either the additive or BM methods. This is due to phonon-induced shifts in the electronic energy levels at strong phonon coupling that have the effect of keeping them out of the bias window for a larger voltage range, and is only correctly captured within the non-additive theory due to the sensitivity of the leads to the augmented (mapped) system. {As in Franck-Condon blockade~\cite{Koch205}, the underlying cause for these shifts is phonon displacement, here captured dynamically through the RC.} Likewise, we find a series of plateaus in the non-additive case for positive bias at lower temperatures in Fig.~\ref{Currentvsbias}(d), as consecutive states of the augmented system enter the bias window with increasing voltage difference \cite{Strasberg16}. In contrast, the leads within the additive and BM theories remain sensitive only to the original DQD eigenstates even at strong phonon coupling, and so exhibit qualitatively similar behaviour to the weak phonon coupling curves.

\begin{figure*}[t]
     \subfloat[ ][Infinite Bias, Lorentzian leads]{{\includegraphics[width=0.45\textwidth]{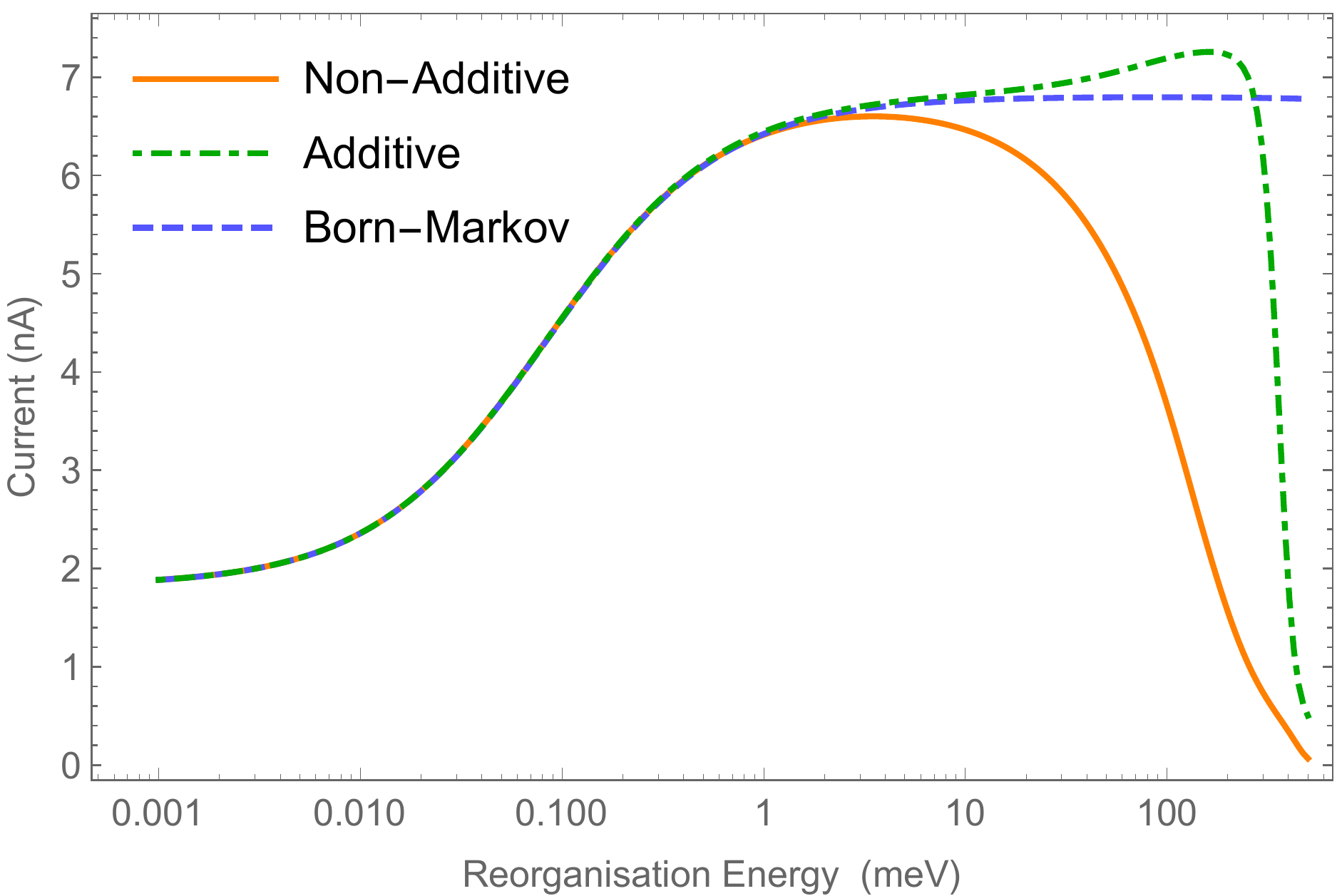}}}%
     \qquad    
     \subfloat[ ][Infinite Bias, Lorentzian leads]{{\includegraphics[width=0.45\textwidth]{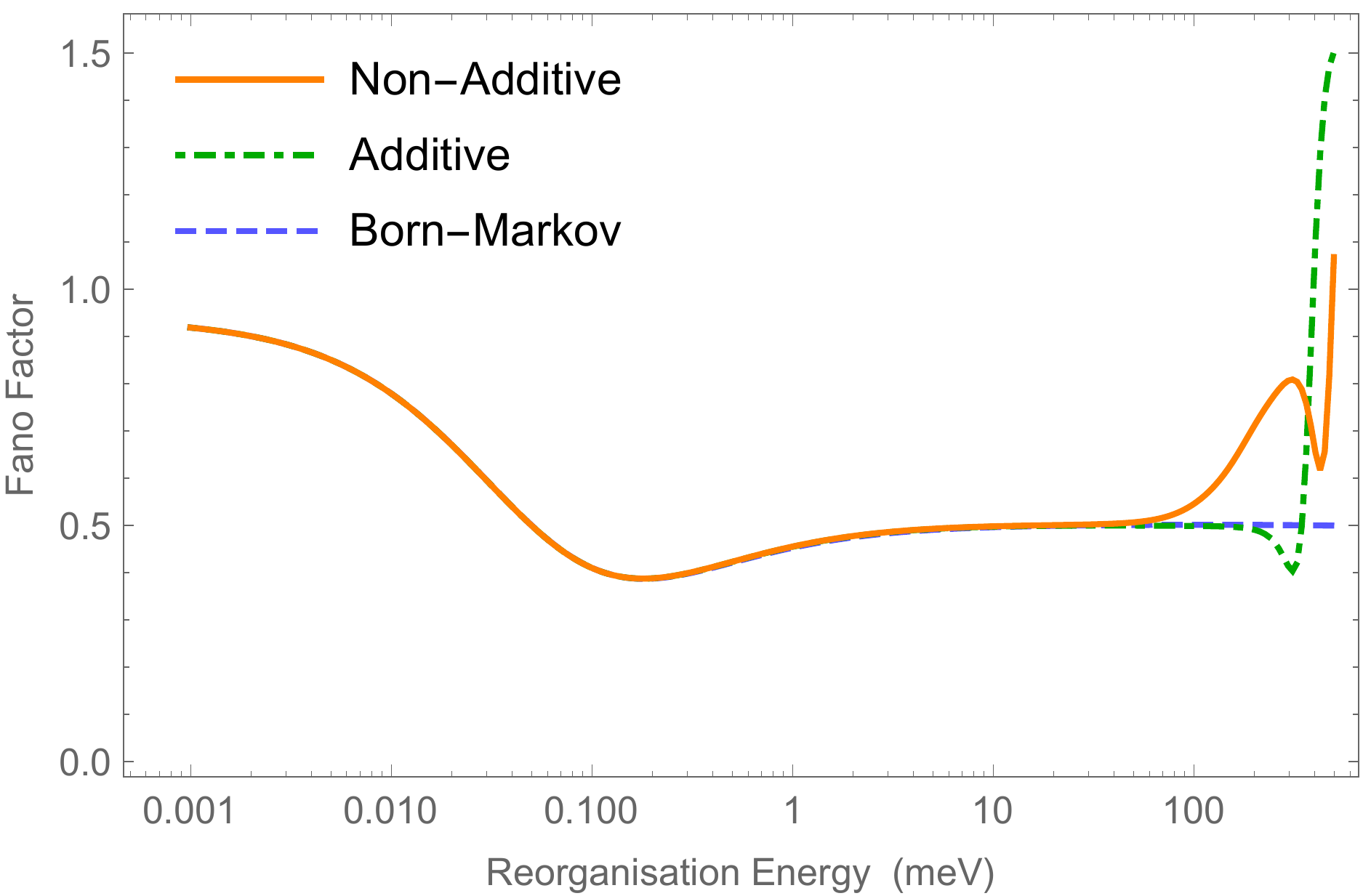}}}%
     \caption{\label{Lorentzian} Current (a) and Fano factor (b) as a function of phonon reorganisation energy $\lambda$ for the case of Lorentzian leads at infinite bias (taken to be $\mu_{\rm L}=-\mu_{\rm R}=10,000$~meV). 
We consider $\delta_{\rm L}=\delta_{\rm R}=100$~{meV} and centre each lead at the energy level of the corresponding dot, $v_{\rm L}=150$~{meV} and $v_{\rm R}=50$~{meV}. Other parameters used are $\epsilon_{\rm L}= 150$~{meV}, $\epsilon_{\rm R}=50$~{meV}, $T=100$~{meV}, $1/\beta_{\rm el}=1/\beta_{\rm ph}=300$~{K}, $\omega_0=\gamma=100$~{meV}, and $\Gamma_{\rm L}= \Gamma_{\rm R} =0.1$~{meV}.}%
\end{figure*}

\subsection{Noise and Fano factor}

{We now move on to look at the impact of non-additive effects on higher order cumulants. We shall focus here on the second order cumulant, which in the case of quantum transport problems is proportional to the zero-frequency noise, as well as its ratio to the first order cumulant, i.e.~the Fano factor. 
Considering first the infinite bias case, in Fig.~\ref{Fanovslambda}(a) we plot the second order cumulant, $\langle\langle I^2\rangle\rangle$, as a function of phonon coupling strength for the same set of parameters as the current in Fig.~\ref{Currentvslambda}. We see that though the overall trends are similar here between  $\langle\langle I^2\rangle\rangle$ 
and current, in the region in which the current begins to increase, $\lambda\approx 0.01-0.1$ meV, the second order cumulant displays a small decline, leading to more ordered transport as indicated by the decrease in Fano factor in Fig.~\ref{Fanovslambda}(b).
As $\lambda$ is increased further $\langle\langle I^2\rangle\rangle$ 
then follows the same trend as the current
as shown by the Fano factor levelling off. 
Finally, at very large $\lambda$, the localisation of the electron onto one of the dots results in a sharp decrease in $\langle\langle I^2\rangle\rangle$.
Nevertheless, for the largest reorganisation energies considered 
the current decreases quicker than $\langle\langle I^2\rangle\rangle$, 
resulting in a transition from a sub-Poissonian to a super-Poissonian Fano factor. As we are working within the Coulomb blockade limit this increased Fano factor does not imply electron bunching within our system, but rather indicates the random nature of the tunneling events \cite{Emary12}. 
Qualitatively similar behaviour is observed at finite bias in Figs.~\ref{Fanovslambda}(c) and (d), though now we see the expected discrepancies between the additive and non-additive theories at large phonon reorganisation energies, with the non-additive Fano factor never becoming super-Poissonian.} 


\subsection{Lorentzian Leads}

So far we have only considered the simplest case of leads with flat spectral densities, using $\Gamma_{\rm L}$ and $\Gamma_{\rm R}$ to denote their couplings to the respective dots. 
We now 
generalise our treatment to Lorentzian lead spectral densities, 
\begin{equation}
\mathcal{J}_{\mathrm{L/R}}(\omega)=\frac{\Gamma _{\mathrm{L/R}}}{2 \pi}\frac{\delta^2_{\mathrm{L/R}}}{(\omega-v_{\mathrm{L/R}})^2+\delta_{\mathrm{L/R}}^2},
\end{equation}
with widths $\delta_{\mathrm{L/R}}$ and peak positions $v_{\mathrm{L/R}}$. This definition results in non-additive contributions being present even in the infinite bias case, as can be seen in Fig.~\ref{Lorentzian}(a) for the current and (b) for the Fano factor. For flat spectral densities at infinite bias the leads become completely insensitive to the eigenstructure of the system Hamiltonian, which results mathematically from the lead correlation functions $C_{12}(\tau)$ and $C_{43}(\tau)$ reducing to delta functions in time ($C_{21}(\tau)=0=C_{34}(\tau)$ in the same limit). This is not true for Lorentzian lead spectral densities. Since the leads are then sensitive to different systems in the additive and non-additive theories, the original S and augmented S$'$, respectively, the agreement between the two that was present at infinite bias for flat leads no longer holds. It is also worth noting from comparison of Figs.~\ref{Currentvslambda}(a) and~\ref{Lorentzian}(a) that the overall current is decreased in the Lorentzian lead case (for the same $\Gamma_L$ and $\Gamma_R$) due to the reduced effective coupling strength away from $\omega=v_{\rm L/R}$. Consequently, phonon coupling effects tend to be slightly more pronounced for Lorentzian leads, given the same set of parameters. 

\section{Conclusions}\label{Conc}

We have employed a combination of the RC mapping and counting statistics techniques to probe electron tunneling through a molecular nanojunction (DQD) system with strong and non-Markovian coupling to phonons. By deriving the lead contributions to the dynamics within the RC mapped (augmented) system representation, we are able to incorporate non-additive phonon-lead effects into the electron counting statistics formalism, which arise here due to the non-negligible phonon interactions.

We find that for flat lead spectral densities, non-additive effects become pronounced for finite lead bias voltages, and that a phenomenological additive approach in which the leads and phonons are treated completely separately can even give rise to unphysical features, such as non-zero current for vanishing lead bias (with all environments held at the same temperature). For Lorentzian leads, non-additive effects manifest themselves even within the infinite bias regime, due to the non-trivial energy dependence of the lead couplings.


In future work it will be interesting to consider the impact of non-additive contributions to the thermoelectric properties of molecular nanojunctions, where we no longer restrict ourselves to regimes in which the phonons and leads are held at the same temperature. 
It would also be intriguing to extend the approach to modelling strong lead couplings, by generalising the RC formalism to fermionic environments~\cite{PhysRevB.97.205405,nazirschaller}. A further avenue of investigation would be to implement counting statistics on the phonon bath to monitor heat flow between the system and the vibrational environment.

\acknowledgments 

We thank Neill Lambert and Gernot Schaller for discussions. C.M. is supported by the EPSRC. A.N. is supported by the EPSRC, grant no. EP/N008154/1.

\appendix

\section{Derivation of Weakly Coupled Leads}\label{AppA}
When considering the lead contributions we 
transform the relevant terms into the system's energy eigenbasis. To do so we diagonalise the system Hamiltonian as given in Eq.~(\ref{HS}). Following the diagonalisation the system Hamiltonian is \begin{equation}\label{HsE}
H_{\rm S}=\Upsilon_1 \ket{\Upsilon_1}\bra{\Upsilon_1} + \Upsilon_2 \ket{\Upsilon_2}\bra{\Upsilon_2}
\end{equation} 
where the $\Upsilon_i$'s are the eigenvalues of $H_{\rm S}$ and the $\ket{\Upsilon_i}$'s are the corresponding eigenvectors, given as
\begin{align}
\Upsilon_1&=\frac{\epsilon_{\rm L}+\epsilon_{\rm R}}{2}+\sqrt{\Delta^2+ T^2},\\
\Upsilon_2&=\frac{\epsilon_{\rm L}+\epsilon_{\rm R}}{2}-\sqrt{\Delta^2+ T^2}, \\
\ket{\Upsilon_1}&=\frac{1}{\sqrt{T^2+\alpha^2}}~\big(\alpha\ket{R}-T\ket{L}\big),\label{E1}\\
\ket{\Upsilon_2}&=\frac{1}{\sqrt{T^2+\alpha^2}}~\big(\alpha\ket{L}+T\ket{R}\big).\label{E2}
\end{align}
Here we have defined $\Delta=\frac{\epsilon_{\rm L}-\epsilon_{\rm R}}{2}$ and $\alpha=\Delta-\sqrt{\Delta^2+T^2}$ for brevity. We also rearrange our eigenvector equations in order to define the original states in terms of the eigenbasis. Rearranging Eqs.~$(\ref{E1})$ and $(\ref{E2})$ we find
\begin{align}
\ket{L}&=\frac{1}{\sqrt{T^2+\alpha^2}}~\big(\alpha\ket{\Upsilon_2}-T\ket{\Upsilon_1}\big)\\
\ket{R}&=\frac{1}{\sqrt{T^2+\alpha^2}}~\big(\alpha\ket{\Upsilon_1}+T\ket{\Upsilon_2}\big).
\end{align}
Having diagonalised the system Hamiltonian we now consider the left lead terms present in the interaction Hamiltonian and express them within the system eigenbasis. Considering Eq.~($\ref{HIL}$) the interaction Hamiltonian is given as
\begin{align}
H_{\rm I}^{\mathrm{L}}&=\frac{1}{\sqrt{T^2+\alpha^2}}~\Big((\alpha\ket{0}\bra{\Upsilon_2}-T\ket{0}\bra{\Upsilon_1})\sum_{k_{\rm L}}t_{k_{\rm L}}c_{k_{\rm L}}^\dagger\nonumber\\&+(\alpha\ket{\Upsilon_2}\bra{0}-T\ket{\Upsilon_1}\bra{0})\sum{k_{\rm L}}t_{k_{\rm L}}^*c_{k_{\rm L}}\Big),
\end{align}
in terms of the system eigenbasis. We define $A_1=\alpha\ket{0}\bra{\Upsilon_2}-T\ket{0}\bra{\Upsilon_1}$ and $A_2=\alpha\ket{\Upsilon_2}\bra{0}-T\ket{\Upsilon_1}\bra{0}$ and find the time dependent form of the two operators in the interaction picture to be
\begin{align}
A_1(t)&=\alpha\ket{0}\bra{\Upsilon_2}e^{-i\Upsilon_2 t}-T\ket{0}\bra{\Upsilon_1}e^{-i\Upsilon_1 t},\\
A_2(t)&=\alpha\ket{\Upsilon_2}\bra{0}e^{i\Upsilon_2 t}-T\ket{\Upsilon_1}\bra{0}e^{i\Upsilon_1 t}.
\end{align}
The environmental contributions are the same as the standard fermionic correlation functions $C_{12}(t)=\sum_{k_{\rm L}}|t_{k_{\rm L}}|^2 e^{i\epsilon_{k_{\rm L}}t} f_{\rm L}(\epsilon_{k_{\rm L}})$ and $C_{21}(t)=\sum_{k_{\rm L}}|t_{k_{\rm L}}|^2 e^{-i\epsilon_{k_{\rm L}}\tau} (1-f_{\rm L}(\epsilon_{k_{\rm L}}))$ \cite{Schaller14}. The Liouvillian for considering only the left lead contributions is then given as
\begin{align}
\mathcal{L}_{\mathrm{L}}\rho_{\rm S}(t)&=-\int^\infty_0 d\tau [A_1,A_2(-\tau)\rho_{\rm S}(t)]C_{12}(\tau)\nonumber\\&+[\rho_{\rm S}(t)A_1(-\tau),A_2]C_{12}(-\tau)\nonumber\\&+[A_2,A_1(-\tau)\rho_{\rm S}(t)]C_{21}(\tau)\nonumber\\&+ [\rho_{\rm S}(t)A_2(-\tau),A_1]C_{21}(-\tau).
\end{align}
After some straightforward algebra we obtain an explicit form for this Liouvillian 
in the energy eigenbasis, which we can express instead in the site basis making use of Eqs.~(\ref{E1}) and (\ref{E2}). We then arrive at the following cumbersome expression for the left lead contribution for a flat spectral density 
\begin{widetext}
\begin{align}\label{LMEEN}
\mathcal{L}_{\mathrm{L}}\rho_{\rm S}(t)&=\frac{\Gamma_{\rm L}}{T^2+\alpha^2}\Big\{f_{\rm L}(\Upsilon_1)\Big(2T^2 \ket{L}\bra{0}\rho_{\rm S}(t) \ket{0}\bra{L}-T\alpha( \ket{L}\bra{0}\rho_{\rm S}(t) \ket{0}\bra{R}+ \ket{R}\bra{0}\rho_{\rm S}(t) \ket{0}\bra{L})\nonumber\\
&-T^2( \ket{0}\bra{0}\rho_{\rm S}(t)+\rho_{\rm S}(t)\ket{0}\bra{0})\Big)+f_{\rm L}(\Upsilon_2)\Big(2\alpha^2 \ket{L}\bra{0}\rho_{\rm S}(t) \ket{0}\bra{L}+T\alpha(\ket{L}\bra{0}\rho_{\rm S}(t) \ket{0}\bra{R}\nonumber\\&+ \ket{R}\bra{0}\rho_{\rm S}(t) \ket{0}\bra{L})
-\alpha^2( \ket{0}\bra{0}\rho_{\rm S}(t)+\rho_{\rm S}(t)\ket{0}\bra{0})\Big)+(1-f_{\rm L}(\Upsilon_1))\Big(2T^2 \ket{0}\bra{L}\rho_{\rm S}(t) \ket{L}\bra{0}\nonumber\\&-T\alpha( \ket{0}\bra{L}\rho_{\rm S}(t) \ket{R}\bra{0}+ \ket{0}\bra{R}\rho_{\rm S}(t) \ket{L}\bra{0})-T^2( \ket{L}\bra{L}\rho_{\rm S}(t)+\rho_{\rm S}(t)\ket{L}\bra{L})\nonumber\\&+T\alpha(\ket{L}\bra{R}\rho_{\rm S}(t)+\rho_{\rm S}(t)\ket{R}\bra{L})\Big)+(1-f_{\rm L}(\Upsilon_2))\Big(2\alpha^2 \ket{0}\bra{L}\rho_{\rm S}(t) \ket{L}\bra{0}\nonumber\\&+T\alpha(\ket{0}\bra{L}\rho_{\rm S}(t) \ket{R}\bra{0}+ \ket{0}\bra{R}\rho_{\rm S}(t) \ket{L}\bra{0})-\alpha^2( \ket{L}\bra{L}\rho_{\rm S}(t)+\rho_{\rm S}(t)\ket{L}\bra{L})\nonumber\\&-T\alpha( \ket{L}\bra{R}\rho_{\rm S}(t)+\rho_{\rm S}(t)\ket{R}\bra{L})\Big)\Big\}.
\end{align}
\end{widetext}
We note that there are no terms that would indicate movement from the ground state to the right dot state (i.e.~terms of the form $ \ket{R}\bra{0}\rho_{\rm S}(t) \ket{0}\bra{R}$) or vice versa, as expected in our DQD set up. We do see, however, that in the finite bias case we have coherence contributions between the left and right dots, 
generated by the inter-dot tunneling parameter, $T$, present in the system Hamiltonian. If we impose the infinite bias limit by taking the Fermi factor $f_{\rm L}\rightarrow 1$ these coherence terms disappear and we are left with 
\begin{align}\label{LIb}
\mathcal{L}_{\mathrm{L}}\rho_{\rm S}(t)^{\mathrm{IB}}&=\Gamma_{\rm L}\big\{\ket{L}\bra{0}\rho_{\rm S}(t)\ket{0}\bra{L}\nonumber\\&-\frac{1}{2} ( \ket{0}\bra{0}\rho_{\rm S}(t)-  \rho_{\rm S}(t)\ket{0}\bra{0})\big\},
\end{align}
which describes hopping of electrons from the left lead to the left dot, rather than to the DQD eigenstates.

The right lead contributions can be stated analogously: 
\begin{widetext}
\begin{align}\label{RMEEN}
\mathcal{L}_{\mathrm{R}}\rho_{\rm S}(t)&=\frac{\Gamma_{\rm R}}{T^2+\alpha^2}\Big\{f_{\rm R}(\Upsilon_1)\Big(2\alpha^2 \ket{R}\bra{0}\rho_{\rm S}(t) \ket{0}\bra{R}-T\alpha(\ket{L}\bra{0}\rho_{\rm S}(t) \ket{0}\bra{R}+ \ket{R}\bra{0}\rho_{\rm S}(t) \ket{0}\bra{L})\nonumber\\
&-\alpha^2( \ket{0}\bra{0}\rho_{\rm S}(t)+\rho_{\rm S}(t)\ket{0}\bra{0})\Big)+f_{\rm R}(\Upsilon_2)\Big(2T^2 \ket{R}\bra{0}\rho_{\rm S}(t) \ket{0}\bra{R}+T\alpha(\ket{L}\bra{0}\rho_{\rm S}(t) \ket{0}\bra{R}\nonumber\\&+ \ket{R}\bra{0}\rho_{\rm S}(t) \ket{0}\bra{L})
-T^2( \ket{0}\bra{0}\rho_{\rm S}(t)+\rho_{\rm S}(t)\ket{0}\bra{0})\Big)+(1-f_{\rm R}(\Upsilon_1))\Big(2\alpha^2 \ket{0}\bra{R}\rho_{\rm S}(t) \ket{R}\bra{0}\nonumber\\&-T\alpha( \ket{0}\bra{L}\rho_{\rm S}(t) \ket{R}\bra{0}+ \ket{0}\bra{R}\rho_{\rm S}(t) \ket{L}\bra{0})-\alpha^2( \ket{R}\bra{R}\rho_{\rm S}(t)+\rho_{\rm S}(t)\ket{R}\bra{R})\nonumber\\&+T\alpha(\ket{R}\bra{L}\rho_{\rm S}(t)+\rho_{\rm S}(t)\ket{L}\bra{R})\Big)+(1-f_{\rm R}(\Upsilon_2))\Big(2T^2 \ket{0}\bra{R}\rho_{\rm S}(t) \ket{R}\bra{0}\nonumber\\&+T\alpha(\ket{0}\bra{L}\rho_{\rm S}(t) \ket{R}\bra{0}+ \ket{0}\bra{R}\rho_{\rm S}(t) \ket{L}\bra{0})-T^2( \ket{R}\bra{R}\rho_{\rm S}(t)+\rho_{\rm S}(t)\ket{R}\bra{R})\nonumber\\&-T\alpha(\ket{R}\bra{L}\rho_{\rm S}(t)+\rho_{\rm S}(t)\ket{L}\bra{R})\Big)\Big\}.
\end{align}
\end{widetext}
As before we can impose the infinite bias limit taking $f_{\rm R}\rightarrow 0$ and we find the Liouvillian reduces to
\begin{align}\label{RIb}
\mathcal{L}_{\mathrm{R}}\rho_{\rm S}(t)^{\mathrm{IB}}&=\Gamma_{\rm R}\big\{\ket{0}\bra{R}\rho_{\rm S}(t)\ket{R}\bra{0}\nonumber\\&-\frac{1}{2}\ket{R}\bra{R}\rho_{\rm S}(t)-\rho_{\rm S}(t)\ket{R}\bra{R}\big\}.
\end{align}
Eqs. (\ref{LIb}) and (\ref{RIb}) are the same as those given, for example, in Ref.~\cite{Stones17} for the lead contributions within the infinite bias regime.

\section{Electron-Phonon Coupling in the RC Formalism}\label{AppB}

When considering molecular nanojunction models we want to be able to model the impact of strong vibrational coupling on the system. In order to do this we make use of the RC mapping to incorporate a collective mode of the vibrational environment within an augmented system. The procedure to do so is discussed here. We ignore the leads for simplicity within this appendix, but their inclusion either additively or non-additively is discussed in the main text.

We begin by considering the post-mapping interaction Hamiltonian of Eq.~(\ref{Hir}). Defining $A_{\mathrm{ph}}=(a^\dagger + a)$, $B_{\mathrm{ph}}=\sum_k f_k(b_k^\dagger + b_k)$ and $\xi = \sum_k \frac{f_k^2}{\nu_k}$ we can write
\begin{equation}
H^\mathrm{ph}_{\rm I'}=A_{\mathrm{ph}}\otimes B_{\mathrm{ph}} + \xi A^2_{\mathrm{ph}}.
\end{equation}
We now derive the associated Liouvillian, beginning with the Liouville-von Neumann equation
\begin{equation}
\dot{\tilde{\rho}}(t)=-i[{H}^{\mathrm{ph}}_{\rm I'}(t),\tilde{\rho}(t)],
\end{equation}
for the full system-environment state $\tilde{\rho}(t)$ within the interaction picture with respect to $H_{\rm S'}+H_{\rm E'}$. 
Treating the residual bath with the BM approximations we arrive at the second-order perturbative form 
\begin{align}\label{neatme}
\tilde{\mathcal L}_{\rm RC}\tilde{\rho}_{\rm S'}(t)&= -i \xi [{A}_{\mathrm{ph}}^2(t),\rho(0)]\nonumber\\&-\xi^2\int^\infty_0 d\tau [{A}_{\mathrm{ph}}^2(t),[{A}_{\mathrm{ph}}^2(t-\tau),\tilde{\rho}_{\rm S'}(t)]]\nonumber\\
&-\int^\infty_0 d\tau\Big( [{A}_{\mathrm{ph}}(t),[{A}_{\mathrm{ph}}(t-\tau),\tilde{\rho}_{\rm S'}(t)]]\Gamma^+(\tau)\nonumber\\&+[{A}_{\mathrm{ph}}(t),\{{A}_{\mathrm{ph}}(t-\tau),\tilde{\rho}_{\rm S'}(t)\}]\Gamma^-(\tau)\Big),
\end{align}
where $\tilde{\rho}_{\rm S'}(t)$ is the reduced system-RC density operator within the interaction picture. 
Here we have defined the bath correlation functions $\Gamma^{\pm}(\tau)=\mathrm{Tr_E}\{(B_{\mathrm{ph}}(\tau)B_{\mathrm{ph}}\pm B_{\mathrm{ph}}(-\tau)B_{\mathrm{ph}})\rho_{\rm E'}\}/2$, with $\rho_{E'}$ a (stationary) thermal state of the residual environment. We can further simplify this Liouvillian by noting that $-i\xi[A_{\mathrm{ph}}^2(t),\rho_{\rm S'}(t)]=-i \xi [A_{\mathrm{ph}}^2(t),\rho(0)]-\xi^2\int^\infty_0 d\tau [A_{\mathrm{ph}}^2(t),[A_{\mathrm{ph}}^2(t-\tau),\rho_{\rm S'}(t)]]$ allowing us to make a substitution that reduces the first two terms into a single commutator. Introducing the forms of the rates $\Gamma^{\pm}$ in the continuum limit,
\begin{align}\label{comms}
\Gamma^+ &=\int^\infty_0 d\omega J_{\rm RC}(\omega)\mathrm{coth}\Big(\frac{\beta \omega}{2}\Big)\mathrm{cos}(\omega \tau),\\ \Gamma^- &=i\int^\infty_0 d\omega J_{\rm RC}(\omega)\mathrm{sin} (\omega \tau),
\end{align}
we are able to rewrite Eq.~(\ref{neatme}) as
\begin{align}\label{nearlydoneme}
\tilde{\mathcal L}_{\rm RC}\rho_{\rm S'}(t)&= -i \xi [A_{\mathrm{ph}}^2(t),\rho_{\rm S'}(t)]\nonumber\\&- \int^\infty_0\int^\infty_0 d\tau d\omega J_{\rm RC}(\omega)\mathrm{coth}\Big(\frac{\beta \omega}{2}\Big)\mathrm{cos}(\omega \tau)\nonumber\\&[A_{\mathrm{ph}}(t),[A_{\mathrm{ph}}(t-\tau),\rho_{\rm S'}(t)]]\nonumber\\& -i\int^\infty_0\int^\infty_0 d\tau d\omega J_{\rm RC}(\omega)\mathrm{sin} (\omega \tau)\nonumber\\&[A_{\mathrm{ph}}(t),\{A_{\mathrm{ph}}(t-\tau),\rho_{\rm S'}(t)\}].
\end{align}
We take the residual bath spectral density to be $J_{\rm RC}(\omega)=\frac{1}{2 \pi}\omega e^{-\omega/ \Lambda}$ and consider the limit of infinite cutoff frequency ($\Lambda \rightarrow \infty$). We then obtain what appear to be divergent terms within our master equation, however it is possible to eliminate these terms by integrating by parts \cite{IlesSmith14,IlesSmith16}. As such we can cancel out the divergences and upon moving back to the Schr\"odinger picture we have 
\begin{align}\label{finalmebeforeRC}
{\mathcal L}_{\rm RC}\rho_{\rm S'}(t)&=-i[H_{\rm S'},\rho_{\rm S'}(t)]\nonumber\\&-\int^\infty_0\int^\infty_0 d\tau d\omega J_{\rm RC}(\omega)\mathrm{coth}\Big(\frac{\beta \omega}{2}\Big)\mathrm{cos}(\omega \tau)\nonumber\\&[A_{\mathrm{ph}},[A_{\mathrm{ph}}(-\tau),\rho_{\rm S'}(t)]]\nonumber\\&-\int^\infty_0\int^\infty_0 d\tau d\omega J_{\rm RC}(\omega)\frac{\mathrm{cos} (\omega \tau)}{\omega}\nonumber\\&[A_{\mathrm{ph}},\{[A_{\mathrm{ph}}(-\tau),H_{\rm S'}],\rho_{\rm S'}(t)\}]
\end{align}
with $H_{\rm S'}$ the mapped system-RC Hamiltonian as before.
We now need to consider our mapped system operators, $A_{\mathrm{ph}}(-\tau)$. To proceed we express them within the augmented system eigenbasis as
\begin{equation}\label{A5eigen}
A_{\mathrm{ph}}=\sum_{jk}A_{\mathrm{ph}_{j,k}}\ket{\psi_j}\bra{\psi_k},
\end{equation}
where the augmented system eigenequation $H_{\rm S'}\ket{\psi_j}=\psi_j \ket{\psi_j}$ is solved numerically, and $A_{\mathrm{ph}_{j,k}}=\bra{\psi_j}A_{\mathrm{ph}}\ket{\psi_k}$. Within the interaction picture with respect to $H_{\rm S'}$ we then have 
\begin{equation}\label{A5teigen}
A_{\mathrm{ph}}(-\tau)=\sum_{jk}A_{\mathrm{ph}_{j,k}}e^{-i\eta_{jk}\tau}\ket{\psi_j}\bra{\psi_k},
\end{equation}
with $\eta_{jk}=\psi_j-\psi_k$ being the eigenvalue differences. Using these definitions we arrive at the final Liouvillian form
\begin{align}
{\mathcal L}_{\rm RC}\rho_{\rm S'}(t)&=-i[H_{\rm S'},\rho_{\rm S'}(t)]-[A_{\mathrm{ph}},[\chi_{\mathrm{ph}},\rho_{\rm S'}(t)]]\nonumber\\&+[A_{\mathrm{ph}},\{\phi_{\mathrm{ph}},\rho_{\rm S'}(t)\}].
\end{align}
Here we have neglected the imaginary Lamb shift terms and included the rates from Eq.~(\ref{finalmebeforeRC}) within the system operators:
\begin{align}
\chi_{\mathrm{ph}}&=\int^\infty_0\int^\infty_0 d\omega d\tau J_{\rm RC}(\omega) \mathrm{cos}(\omega \tau)\mathrm{coth}\Big(\frac{\beta \omega}{2}\Big) A_{\mathrm{ph}}(-\tau)\nonumber \\
&\approx \frac{\pi}{2} \sum_{j k} J_{\rm RC}(\eta_{jk})\mathrm{coth}\Big(\frac{\beta \eta_{jk}}{2}\Big) A_{\mathrm{ph}_{jk}}\ket{\psi_j}\bra{\psi_k},\\
\phi_{\mathrm{ph}}&=\int^\infty_0\int^\infty_0 d\omega d\tau \frac{J_{\rm RC}(\omega) \mathrm{cos}(\omega \tau)}{\omega} [H_{\rm S'},A_{\mathrm{ph}}(-\tau)]\nonumber\\
&\approx \frac{\pi}{2} \sum_{j k} J_{\rm RC}(\eta_{jk}) A_{\mathrm{ph}_{jk}}\ket{\psi_j}\bra{\psi_k}.
\end{align}

%

\end{document}